\documentclass[pre, twocolumn,superscriptaddress, amsmath,amssymb,aps]{revtex4-2}

\usepackage{graphicx}
\usepackage{dcolumn}
\usepackage{bm}
\usepackage{svg} 
\usepackage{MnSymbol}
\usepackage{svg}
\usepackage{stmaryrd}
\usepackage[capitalise]{cleveref} 

\begin{document}

\title{Topologically-constrained fluctuations and thermodynamics regulate nonequilibrium response}

\author{Gabriela Fernandes Martins}
\email{gmartins@umich.edu}
\affiliation{Department of Physics, University of Michigan, Ann Arbor, Michigan 48109, USA}

\author{Jordan M. Horowitz}
\email{jmhorow@umich.edu}
\affiliation{Department of Biophysics, University of Michigan, Ann Arbor, Michigan, 48109, USA}
\affiliation{Center for the Study of Complex Systems, University of Michigan, Ann Arbor, Michigan 48104, USA}
\affiliation{Department of Physics, University of Michigan, Ann Arbor, Michigan 48109, USA}

\begin{abstract}
Limits on a system's response to external perturbations inform our understanding of how physical properties can be shaped by microscopic characteristics.
Here, we derive constraints on the steady-state nonequilibrium response of physical observables in terms of the topology of the microscopic state space and the strength of thermodynamic driving.
Notably, evaluation of these limits requires no kinetic information beyond the state-space structure.
When applied to models of receptor binding, we find that sensitivity is bounded by the steepness of a Hill function with a Hill coefficient enhanced by the chemical driving beyond the structural equilibrium limit.

\end{abstract}

\maketitle

\section{Introduction}

A useful method for understanding physical properties of a system in and out of equilibrium is to analyze how it responds to external perturbations~\cite{Kubo}. For example, material coefficients, like diffusivity  and viscoelasticity, are basic inputs into any soft matter description~\cite{Chaikin,Mason1995,Mizuno2007,BenIsaac2011,Fakhri2014,Chun2021,Bowick2022}. Another example is how sensitivity to chemical inputs is used as a key performance measure for a variety of biophysical processes, from biochemical sensing~\cite{Bialek2005,Govern2014b} to gene transcription~\cite{Lan2012,Sartori2015} and beyond~\cite{Murugan2014,Hartich2015,Wong2017,Cui2018,Mallory2019,Estrada2016,Goldbeter1981,Qian2003,Owen2022,Tran2018,Park2019,Reimer2023,Kim2022}.

When the system is near equilibrium the Fluctuation-Dissipation Theorem (FDT) operates as a powerful organizing principle~\cite{Kubo}: fluctuations and response encode the same information. The FDT's utility has led to significant interest in developing similar predictions valid far from equilibrium~\cite{Harada2005,Lippiello2014,Wang2016,Nardini2017,Dadhichi2018,Baiesi2013,Baldovin2022}. Some link the response to fluctuations in particular physical observables~\cite{Agarwal1972,Prost2009,Seifert2010,Speck2006,Baiesi2009,Chetrite2009,Seifert2010b,Chaudhuri2012,Bohec2013,Caprini2021} while others restrict attention to specific equilibrium-like perturbations~\cite{Graham1977,Lubensky2010,Chun2021} or preparations~\cite{Altaner2016}.

In recent years an alternative approach has emerged, where trade-offs or inequalities delineate the limits of possible behavior~\cite{Uhl2019,Dechant2020,Baiesi2011}. One such class of predictions are the thermodynamic uncertainty relations, which are thermodynamic and kinetic bounds on fluctuations~\cite{Barato2015,Gingrich2016,Horowitz2020,PelitiST,Barato2017}. Here, we build on another class of trade-offs, a recently established collection of thermodynamic bounds on steady-state response~\cite{Owen2020,Owen2023,Gao2021}. These past predictions are limited by not accounting for correlations between responses at different microscopic configurations. Here we include these correlations and demonstrate that the response of a physical observable is bounded not just by thermodynamic driving, but also by a measure of fluctuations sensitive to the topology of the microscopic state space. Importantly, the only kinetic information required to determine these fluctuations is the structure of the state-space; no knowledge of the values of kinetic rates is needed.

Our theoretical tools are graph-theoretic solutions to the steady-state distribution and its derivatives with respect to kinetic rates. Such representations have been known for some time~\cite{schnakenberg1976,Hill}, and in recent years have re-emerged as powerful tools for studying the links between kinetics and thermodynamics in noisy nonequilibrium systems~\cite{Andrieux2006,Maes2012,Polettini2017,Khodabandehlou2022,Liang2022,Nam2022,Cetiner2022}. 

\section{Dynamics and thermodynamics}
\begin{figure}
\includegraphics[width=8.6cm]{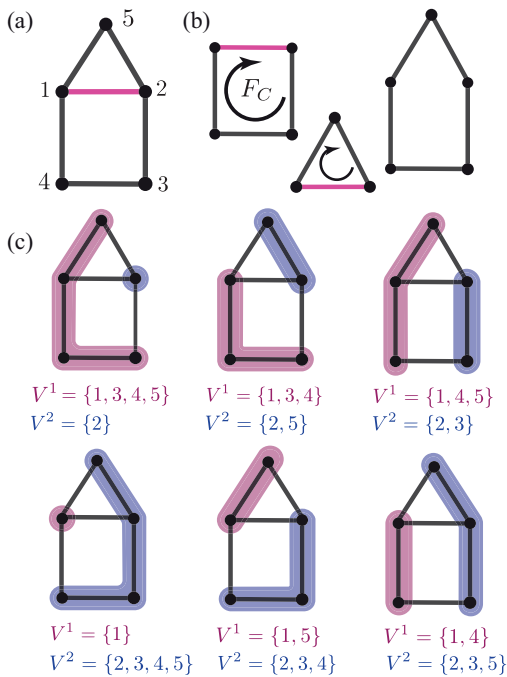}
\caption{Representative example: (a) Five-state house graph with perturbed edge $e_{12}$ highlighted in pink. (b) Cycles associated to the thermodynamic forces $F_C$.  Only cycles through the perturbed edge $e_{12}$ constrain the response. (c) Topologically-consistent splittings ${\mathcal V}^{12}$ with $V^1$ highlighted in pink and $V^2$ in blue.}
\label{fig:graph}
\end{figure}

Consider a system whose dynamics can be modeled as a Markov jump process, making random transitions among a collection of states, or configurations, $i=1,\dots, N$ with rates $W_{ij}$ to jump from $j\to i$.
For thermodynamic consistency~\cite{PelitiST}, we will assume every transition is accompanied by its reverse ($W_{ji}\neq 0$ whenever $W_{ij}\neq 0$).
We can then visualize these dynamics occurring on a state-space graph, $G$, where the vertices $\left\{i\right\}$ represent the states and the (undirected) edges $\{e_{mn}\}$ represent allowed transitions in both directions.
An example which will serve to illustrate our results is introduced in Fig.~\ref{fig:graph}(a).

For such models, the probability $p_i(t)$ for the system to be at state $i$ at time $t$ evolves according to the master equation~\cite{VanKampen}
\begin{equation}\label{eq:master} 
\dot{p}_{i}(t)=\sum_{j=1}^{N}W_{ij}p_{j}(t),
\end{equation}
where the  elements $W_{ii}=-\sum_{j\neq i} W_{ji}$ have been introduced to enforce probability conservation.
We will assume that the graph $G$ is (strongly) connected, which coupled with our assumption that every transition has a reverse, guarantees that $p_i(t)$ relaxes to a unique stationary distribution $\pi_i$ given as the solution of $\sum_j W_{ij}\pi_j=0$.

Driven, or nonequilibrium dynamics are characterized by the emergence of nonzero steady-state probability currents between pairs of states, ${\bar J}_{ij} = W_{ij}\pi_j-W_{ji}\pi_i$.
These flows are driven by thermodynamic forces---like temperature or chemical potential differences---and are linked to the dynamics by assuming local detailed balance~\cite{Seifert2012,VandenBroeck2015}.
This allows us to identify these forces through the imbalances of the rates around cycles: sequences of edges that connect the initial vertex to itself without self-intersection (Fig.~\ref{fig:graph}(b)).
Then to each cycle $C=\left\{i_0\rightarrow i_1\rightarrow i_2\rightarrow\cdots\,\, i_m\rightarrow i_0\right\}$ we identify a thermodynamic force, or affinity, as the log-ratio of rates forwards and backwards around the cycle~\cite{schnakenberg1976,Andrieux2006,PelitiST}:
\begin{equation}\label{eq:cycle_force}
F_{C}=\ln\left(\frac{W_{i_0i_m}\ldots W_{i_2i_1}W_{i_1i_0}}{W_{i_0i_1}W_{i_1i_2}\ldots W_{i_mi_0}}\right).
\end{equation}
When all cycle forces vanish, then necessarily all steady-state currents are zero (${\bar J}_{ij}=0$), and the system satisfies detailed balance, which is a statistical symmetry characteristic of equilibrium.
Put another way, the larger the cycle forces, the farther from equilibrium the steady state.

\section{Response}
\label{response}

Typically, the experimentally accessible quantity is not the steady-state distribution, but steady-state averages of observables $\langle Q\rangle = \sum_i Q_i \pi_i$.
When the steady state is detailed balanced we will use the superscript `${\rm eq}$' to distinguish such equilibrium averages.
It is then our goal to predict how perturbations of the rates affect these averages.

We model perturbations by allowing the rates to depend on an externally-controlled parameter $\lambda$.
A common, physically-motivated choice for this dependence is to exponentially re-weight the rates $W_{ij}(\lambda)=W_{ij}e^{\lambda d_{ij}}$ through a coupling $d_{ij}$ for $i\neq j$, which may have nonzero symmetric $d^s_{ij}=d_{ij}+d_{ji}$ and asymmetric $d^a_{ij}=d_{ij}-d_{ji}$ parts~\cite{Baiesi2013}.
In which case, the steady-state (or static) response to an external perturbation is defined by the linear combination of logarithmic derivatives
\begin{equation}\label{eq:derivative}
\begin{split}
\frac{\partial \langle Q\rangle}{\partial \lambda} &= \sum_{m\neq n} d_{mn}W_{mn}\frac{\partial\langle Q\rangle}{\partial W_{mn}} \\
&= \sum_{i,m\neq n} Q_i d_{mn}W_{mn}\frac{\partial \pi_i}{\partial W_{mn}}. 
\end{split}
\end{equation}

\subsection{Equilibrium steady states}

For perturbations around equilibrium steady-states $\pi^{\rm eq}$ where ${\bar J}^{\rm eq}_{mn}=0$, the FDT links the equilibrium response to the fluctuations~\cite{PelitiST,Seifert2012}:
\begin{equation}\label{eq:FDT}
\frac{\partial \langle Q\rangle^{\rm eq}}{\partial \lambda} =  \int_0^\infty  \langle Q(t) J_{d^a}(0)\rangle^{\rm eq} dt,
\end{equation}
 where the two-time correlation function between the observable and the $d$-weighted current is defined as $\langle Q(t) J_{d^a}(0)\rangle= \sum_{lij}Q_l(e^{tW})_{li}d^{a}_{ij}W_{ij}\pi_j$ in terms of the transition probability $p(l,t|i,0)=(e^{tW})_{li}$ for the system to be at $l$ at time $t$ given it was initially at $i$.
Importantly, only the asymmetric part of the coupling $d^a_{ij}$ contributes.
The symmetric part $d^s_{ij}$ amounts to a coordinated and equal change in the forward and reverse rates between a pair states. 
It is akin to varying a kinetic barrier, which cannot alter a system at equilibrium---the equilibrium Gibbs distribution only depends on the energies and not on the kinetics.
It is worth noting that generically adding an asymmetric coupling induces a nonconservative force, driving the system slightly away from equilibrium.  
The exception is when the (asymmetric) coupling is  derivable from a potential, $d^a_{ij}  = -(U_i-U_j)$: in this case $J_d = -{\dot U}$, and the FDT \eqref{eq:FDT} simplifies to a static equilibrium correlation,
\begin{equation}\label{eq:FDT_ener}
\frac{\partial \langle Q\rangle^{\rm eq}}{\partial \lambda} = -  \int_0^\infty  \langle Q(t) {\dot U}(0)\rangle^{\rm eq} dt = \left\llangle Q,U\right\rrangle^{\rm eq},
\end{equation}
where the covariance is $\left\llangle Q,U\right\rrangle=\langle QU\rangle-\langle Q\rangle\langle U\rangle$.
This perturbation is tantamount to varying the system's energy landscape by including a new potential $U_i$.

\subsection{Nonequilibrium steady states}

In light of our discussion, it is natural when studying nonequilibrium response to individually address changes in symmetric, asymmetric, or other particular combinations of rates.
Our previous work has identified the following combinations of logarithmic derivatives as useful~\cite{Owen2020}
\begin{align}
\label{eq:decomposition_E}
& \frac{\partial }{\partial E_i} = -\sum_{j\neq i}W_{ji}\frac{\partial}{\partial W_{ji}}\\
   & \frac{\partial }{\partial B_{mn}}=-W_{mn}\frac{\partial}{\partial W_{mn}}-W_{nm}\frac{\partial}{\partial W_{nm}}\label{eq:decomposition_B} \\ 
   & \frac{\partial }{\partial F_{mn}}=\frac{1}{2} \left(W_{mn}\frac{\partial}{\partial W_{mn}}-W_{nm}\frac{\partial}{\partial W_{nm}}\right) \label{eq:decomposition_F}.
\end{align}

The $E$-perturbations \eqref{eq:decomposition_E}, which are uniform changes in the total exit rate from a state, are energy-like or equilibrium-like in that they satisfy a fluctuation-response equality akin to the equilibrium FDT \eqref{eq:FDT_ener}, but valid arbitrarily far from equilibrium~\cite{Owen2020}
\begin{equation}\label{eq:E-perturbation}
\sum_{i=1}^N U_i \frac{\partial \langle Q\rangle}{\partial E_i} = \left\llangle Q,U\right\rrangle,
\end{equation}
for arbitrary state function $U_i$.
This prediction holds more generally, applying to diffusion processes  as well as  time-dependent response~\cite{Graham1977,Chun2021}.

The symmetric $B$-perturbations \eqref{eq:decomposition_B} are like changes in kinetic `barriers', and the asymmetric $F$-perturbations~\eqref{eq:decomposition_F} are like shifts in the driving forces.
For these perturbations, we previously demonstrated constraints on ratios of nonnegative observables, $Q^{(1)}, Q^{(2)}\ge 0$~\cite{Owen2020}:
\begin{align}\label{eq:oldB}
\left| \frac{\partial \ln(\langle Q^{(1)}\rangle/\langle Q^{(2)}\rangle)}{\partial B_{mn}}\right| &\le \tanh({\mathcal F}_{\rm max}/4) \\ 
\label{eq:oldF}
\left| \frac{\partial \ln(\langle Q^{(1)}\rangle/\langle Q^{(2)}\rangle)}{\partial F_{mn}}\right|&\le 1,
  \end{align}
where ${\mathcal F}_{\rm max}=\max_{C\ni e_{mn}} F_C$ is the maximum cycle force through the perturbed edge.

Our focus here is the response of a single observable $\langle Q\rangle$ \eqref{eq:derivative}, not a ratio.
To transform the predictions in \eqref{eq:oldB} and \eqref{eq:oldF} into bounds on a single observable, let us introduce notation for the observable's maximum $Q_M$ and minimum $Q_m$.
Then, by setting the two positive observables in  \eqref{eq:oldB} and \eqref{eq:oldF} to be $Q^{(1)} = Q-Q_m\ge 0$ and $Q^{(2)}=Q_M-Q\ge 0$, we arrive at the relevant predictions
\begin{align}\label{eq:oldB2}
\left| \frac{\partial \langle Q\rangle}{\partial B_{mn}}\right| &\le \frac{\left(Q_{M}-\langle Q\rangle\right)\left(\langle Q\rangle - Q_{m}\right)}{Q_{M}-Q_{m}} \tanh({\mathcal F}_{\rm max}/4) \\ 
\label{eq:oldF2}
\left| \frac{\partial \langle Q\rangle}{\partial F_{mn}}\right| &\le \frac{\left(Q_{M}-\langle Q\rangle\right)\left(\langle Q\rangle - Q_{m}\right)}{Q_{M}-Q_{m}}.
  \end{align}
These predictions do not account for any specific properties of the network's topology.
In the following, we provide tighter inequalities that reveal how the topology of $G$ interfaces with the thermodynamics to limit nonequilibrium response.

\section{Symmetric perturbations} 
\subsection{Single edge}
We begin our analysis by determining the maximum response to a symmetric perturbation along a single edge $e_{mn}$.
As we noted, symmetric perturbations cannot generate any response at equilibrium.
So nonequilibrium driving is required, and as we will show quantitatively bounds the response.
Here we summarize the derivation, details can be found in Appendices~\ref{sec:Methods} and ~\ref{sec:symmetric}.

To proceed, we differentiate the master equation at steady-state \eqref{eq:master} to obtain a set of inhomogenous linear equations for the responses of the steady-state distribution,
\begin{equation}\label{eq:nonhomo_B}
    \sum_{j=1}^N W_{ij}\frac{\partial \pi_j}{\partial{B_{mn}}}={\bar J}_{mn}\left(\delta_{im}-\delta_{in}\right).
\end{equation}
Our main theoretical tool is then a graph-theoretic solution to this set of equations in terms of spanning two-forests of $G$, which was originally derived in~\cite{Caplan1982}, though we require a slight modification presented in Appendix~\ref{sec:Methods}.
Substitution of this graphical representation into \eqref{eq:derivative}, allows us to reformulate the question of bounding the response as a linear  optimization problem.
The optima then serve as potential upper bounds.
The form of these optima is inherited from our graph-theoretic analysis and therefore depends on the topology of the state space.
The required quantity we call a topologically-consistent splitting of the vertices of the graph, $V^{mn}\in {\mathcal V}^{mn}$: each $V^{mn}$ is formed by cutting the graph 
 into two connected components (that are disjoint), one  $V^m$ which contains the vertex $m$ and the other $V^n$ which contains vertex $n$.
We then denote the indicator function on the states in one of these components as $\delta_i(V^m)$, taking the value one when $i\in V^m$ and zero otherwise.

Our first main result is that topology and thermodynamics constrain the maximum response via
\begin{equation}\label{eq:bound_B}
    \left|\frac{\partial \langle Q \rangle}{\partial B_{mn}}\right|\leq\underset{\mathcal{V}^{mn}}{\max} \Big|\left\llangle Q,\delta(V^{m})\right\rrangle\Big| \tanh\left({\mathcal F}_{\rm max}/4\right),
\end{equation}
with ${\mathcal F}_{\rm max}=\max_{C\ni e_{mn}} F_C$.
Note that placing the indicator function on $\delta(V^m)$ is equivalent to $\delta(V^n)=1-\delta(V^m)$ due to the linearity of the covariance  
 and $\llangle Q,1\rrangle = 0$: $\left\llangle Q,\delta(V^{m})\right\rrangle=-\left\llangle Q,\delta(V^{n})\right\rrangle$.

\begin{figure}[t]
\includegraphics[width=8.6cm]{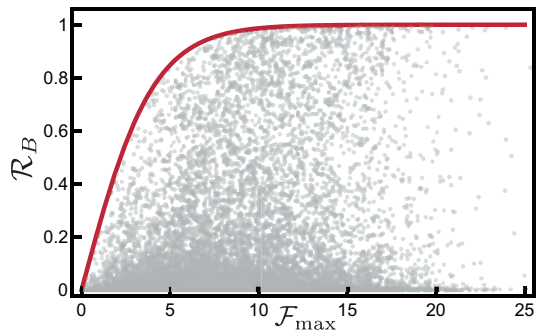}
\caption{Response ratio $\mathcal{R}_B$ as a function of the maximum cycle force ${\mathcal F}_{\rm max}$ around cycles containing edge $e_{12}$ of the house graph (Fig.~\ref{fig:graph}) for 15000 rate matrices with the logarithmic asymmetric $\ln(W_{ij}/W_{ji})$ and symmetric $\ln(W_{ij} W_{ji})$ parts sampled uniformly on $[-7,7]$ and the observable $Q_i$ sampled uniformly on $[-4,4]$. All samples (gray) fall below the predicted bound $\tanh (\mathcal{F}_{\rm max}/4)$ (red).}
\label{fig:bound}
\end{figure}

Equation \eqref{eq:bound_B} has the character of the FDT, linking response to fluctuations.
Here, however, the covariance measures the fluctuations of the observable across the two components of $V^{mn}$, a topologically-dependent noise characteristic.
The response is zero whenever all the cycle forces through the perturbed edge are zero (${\mathcal F}_{\rm max}=0$).
This is possible away from equilibrium, but is always true at equilibrium.
It also can occur if there are no cycles through the perturbed edge because it is a bridge---its removal disconnects $G$ into two disjoint components.
In Fig.~\ref{fig:bound}, we verify \eqref{eq:bound_B} by plotting the response ratio ${\mathcal R}_B = |\partial_{B_{mn}}\langle Q\rangle| /\max_{\mathcal{V}^{mn}} |\left\llangle Q,\delta(V^{m})\right\rrangle|$ for perturbations along $e_{12}$ of the graph in Fig.~\ref{fig:graph}(a) for random observables and rates. 

\subsection{Multiple edges}
\begin{figure}[b!]
\includegraphics[width=8.6cm]{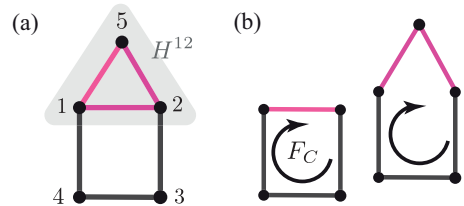}
\caption{(a) Representation of a multi-edge perturbation for the illustrative graph. All edges in the subgraph $H^{12}$, highlighted in gray, are uniformly perturbed. (b) Cycles that count towards determining $\mathcal{F}_{\max}$ in the case of multi-edge perturbation of $H^{12}$. Note that the cycle $\{1 \to 2 \to 5 \to 1\}$ is not relevant in this case, as it is internal to $H^{12}$.}
\label{fig:multi}
\end{figure}

When multiple edges are perturbed in the network, we expect the response to have a more complicated dependence on the topology.  
However, there is one situation where our analysis directly generalizes to multi-edge perturbations.
That is when their combined impact is effectively like a single edge perturbation.
Here, we have in mind the situation illustrated in Fig.~\ref{fig:multi} where we perturb a set of edges in a subgraph $H^{mn}$ of $G$ that only connects to the rest of the graph at two vertices, which we will call $m$ and $n$, with a slight abuse of notation. In this case our analysis carries through with minimal as described in Appendix~\ref{sec:mutli-symmetric}, with the result
 \begin{equation}\label{eq:bound_B_multi}
    \left|\sum_{e_{kl}\in H^{mn}}\frac{\partial \langle Q \rangle}{\partial B_{kl}}\right|\leq\underset{\mathcal{V}^{mn}}{\max} \Big|\left\llangle Q,\delta(V^{m})\right\rrangle\Big| \tanh\left({\mathcal F}_{\rm max}/4\right).
\end{equation}
Here, ${\mathcal F}_{\rm max}$ is the maximum cycle force over all cycles that straddle the perturbed and unperturbed regions passing through the two vertices $m$ and $n$.

\subsection{Design principles and optimal topologies}

Analysis of the derivation of \eqref{eq:bound_B} allows us to identify design principles for achieving the maximum response by determining how we should tune the rates $W_{ij}$ to saturate the inequality.
By changing the rates, we can effectively change the network topology of the state-space graph, thus identifying what effective network structures are optimal.

There are two limits of the rates that we will encounter that lead to effective changes in the network structure: sending the rates to zero or to infinity.
If we drive a pair of rates along a single edge $e_{ij}$ to zero ($W_{ij}=W_{ji}=0$) the resulting dynamics take place on a state-space graph with that edge deleted, as transitions along that edge are no longer possible. 
On the other hand, if we take a pair of rates on an edge $e_{ij}$ to be large, the system will relax to a local steady state on the pair of nodes $i$ and $j$ very quickly: the remaining slow dynamics evolve on a state-space graph where the nodes $i$ and $j$ have been contracted into a single node.

\begin{figure}[b!]
\includegraphics[width=8.6cm]{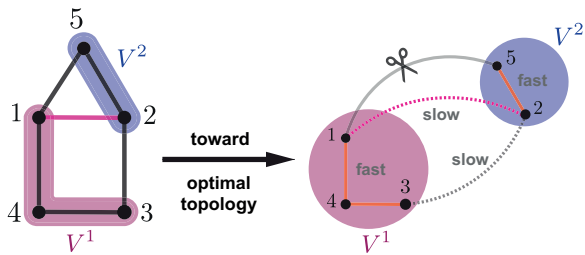}
\caption{Example of an optimal network topology inherited from the topologically-consistent splitting $V^1=\{1,3,4\}$, $V^2=\{2,5\}$. The rates on edge $e_{15}$ are set to zero cutting that edge from the graph, leaving the system with a unique cycle. States internal to the islands defined by the topological splitting are contracted into a pair of single nodes by taking the rates along edges internal to the islands (orange)  to be fast compared to the rates connecting the islands (dashed).}
\label{fig:opt_network}
\end{figure}

We have found that to saturate \eqref{eq:bound_B} two conditions are required.
The first is that there is a single cycle passing through our states $m$ and $n$.
Thus, we have to interrupt all cycles but one through $m$ and $n$ by deleting at least one of their edges (without disconnecting the graph) by sending the rates along that edge to zero.
The second condition is that the graph has a single dominant topologically-consistent splitting $V^{mn}$.
This is accomplished by taking all the rates on each vertex set $V^m$ and $V^n$ to be large such that they each form isolated dynamical islands.
These islands are then linked together by a pair of slow rates that complete the unique cycle in the system.
For the house graph, this is illustrated in Fig.~\ref{fig:opt_network} for a particular topologically-consistent splitting.
The emergent optimal network acts as a two state system, linked up by a single cycle.

\section{Asymmetric and arbitrary perturbations} 
Nonsymmetric edge perturbations can generate a response even in equilibrium~\eqref{eq:FDT}.  Thus, we expect thermodynamics to not be a limiting constraint, and focus solely on the influence of network topology.  We first will analyze a general rate perturbation, and then specialize to asymmetric perturbations~\eqref{eq:decomposition_F}.
Details can be found in Appendix~\ref{sec:single_rate}.

Again, we differentiate the master equation \eqref{eq:master} to obtain a set of inhomogenous linear equations for the responses to logarithmic-perturbations in a single rate constant $W_{mn}$,
\begin{equation}\label{eq:nonhomo_F}
    \sum_{j=1}^N W_{ij}\frac{\partial \pi_j}{\partial{\ln W_{mn}}}=W_{mn}\pi_n\left(\delta_{in}-\delta_{im}\right).
\end{equation}
Utilizing our graph-theoretic representation leads to a linear optimization problem.
Its solution gives an identical topological bound for a rate perturbation
\begin{equation}\label{eq:W_perturb}
    \left|\frac{\partial \langle Q \rangle}{\partial \ln W_{mn}}\right|\leq\underset{\mathcal{V}^{mn}}{\max} \Big|\left\llangle Q,\delta(V^{m})\right\rrangle\Big|.
\end{equation}
We see the correlation between the observable and the topologically-consistent splittings provide the ultimate limit, no matter how strongly driven the system.

Equation \eqref{eq:W_perturb} readily leads to a constraint on asymmetric perturbations as well,
\begin{equation}
\begin{split}
    \left|\frac{\partial \langle Q \rangle}{\partial F_{mn}}\right|&=\frac{1}{2}\left|\frac{\partial \langle Q \rangle}{\partial \ln W_{mn}}-\frac{\partial \langle Q \rangle}{\partial \ln W_{nm}}\right| \\
    &\le\underset{\mathcal{V}^{mn}}{\max} \Big|\left\llangle Q,\delta(V^{m})\right\rrangle\Big|,
    \end{split}
\end{equation}
after noting that the covariance bound is symmetric with respect to $W_{mn}$ and $W_{nm}$.

This inequality is particular interesting when applied to asymmetric perturbations around equilibrium steady-states.
In this case, we can apply the FDT \eqref{eq:FDT} to get a nontrivial bound on the two-time correlation function between any observable and the current $J_{mn}$ ($d^a_{ij} = \delta_{mi}\delta_{nj}$) on the perturbed edge
\begin{equation}
    \left|\int_0^\infty  \langle Q(t) J_{mn}(0)\rangle^{\rm eq} dt\right|\leq\underset{\mathcal{V}^{mn}}{\max} \Big|\left\llangle Q,\delta(V^{m})\right\rrangle^{\rm eq}\Big|.
\end{equation}
Structure quantifiably constrains fluctuations near equilibrium as well.

\section{Operational limits}
Each of our bounds depends on a covariance with the indicator function $\delta(V^m)$.
Measuring such correlations requires access to the occupation statistics of the states.  
This may be challenging even for moderately sized systems, where only coarser observations are possible. 
Thus, we now turn to deriving weakened, operational bounds on the covariance that depend only on observable properties of $Q$: its average $\langle Q\rangle$, variance $\llangle Q^2\rrangle$, maximum $Q_M$, and minimum $Q_m$.

To bound the covariance, we note that each $\delta(V^m)$ is nonnegative and bounded by one ($0\le \delta(V^m)\le 1$).
Let us consider the set of all such bounded observables, ${\mathcal BO} = \{A| 0\le A_i\le 1\}$, of which $\delta(V^m)$ is a member.
Then, the correlation we wish to constrain can trivially be bounded by the maximum over all steady-states $\pi_i$ and all bounded observables, keeping the average $\langle Q\rangle$ and variance $\llangle Q^2\rrangle$ fixed:
\begin{equation}
\underset{\mathcal{V}^{mn}}{\max} \Big|\left\llangle Q,\delta(V^{m})\right\rrangle\Big| \le \underset{\{\pi,A\in {\mathcal BO}|\langle Q\rangle, \llangle Q^2\rrangle \}}{\max} \Big|\left\llangle Q,A\right\rrangle\Big|.
\end{equation}
This weaker optimization can be carried out (Appendix~\ref{sec:opt_limits}), with a result that depends on $\Gamma_Q = \min \{Q_{M}-\langle Q\rangle,\langle Q\rangle - Q_{m},\sqrt{\llangle Q^2\rrangle}\}$ via
\begin{equation}\label{eq:operational3}
\underset{\mathcal{V}^{mn}}{\max} \Big|\left\llangle Q,\delta(V^{m})\right\rrangle\Big| \le \frac{\Gamma_Q \llangle Q^2\rrangle}{\Gamma_Q^2+\llangle Q^2\rrangle},
\end{equation}
where the variance is required to fall below $\llangle Q^2\rrangle \le \left(Q_{M}-\langle Q\rangle\right)\left(\langle Q\rangle - Q_{m}\right)$.
 Saturation occurs when $\pi_i$ is nonzero on only two states whose precise values depend on the value of $\Gamma_Q$.

To make contact with the bounds \eqref{eq:oldB2} and \eqref{eq:oldF2} derived in \cite{Owen2020}, we relax the constraint on the variance.
Observing that \eqref{eq:operational3} is monotonically increasing function of the variance, we can bound it  by setting the variance to its maximum value  $\llangle Q^2\rrangle = \left(Q_{M}-\langle Q\rangle\right)\left(\langle Q\rangle - Q_{m}\right)$:
\begin{align}\label{eq:operational}
\underset{\mathcal{V}^{mn}}{\max} \Big|\left\llangle Q,\delta(V^{m})\right\rrangle\Big| &\le \frac{\left(Q_{M}-\langle Q\rangle\right)\left(\langle Q\rangle - Q_{m}\right)}{Q_{M}-Q_{m}}\\
\label{eq:operational2}
&\le  \frac{1}{4}\left(Q_{M}-Q_{m}\right),
\end{align}
where in the second line we have further maximized over all $Q_m\le \langle Q\rangle \le Q_M$.
The first bound \eqref{eq:operational} is saturated when $\pi_i$ is nonzero on only two states: one of the states, call it $i=M$, where $Q_i$ reaches its maximum value $Q_M$ and another state $i=m$ where the observable reaches its minimum value $Q_m$.  
The weaker bound \eqref{eq:operational2} saturates when the probability is evenly split between those two states, $\pi_M=\pi_m=1/2$.

By arriving at the previously derived bounds \eqref{eq:oldB2} and \eqref{eq:oldF2} through our more refined inequalities, we uncover how they emerge as the maximum response over all steady-state distributions with the mean observable held fixed.
Under these conditions, the steady-state distribution is peaked at only two states.
Thus, our more refined bounds provide the limits to steady-state response for an arbitrary steady-state distribution that can be spread among multiple states, accounting for how the response at different states must be related through topology as manifested through fluctuations across the topologically-consistent splittings.
One could further imagine tighter operational bounds that constrain additional cumulants of the observables $Q$ and $\delta(V^m)$.

\section{Illustrative example: receptor binding}
\begin{figure}
\includegraphics[width=8.6cm]{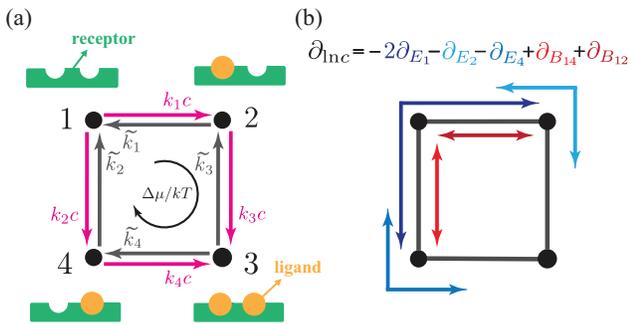}
\caption{Receptor binding: (a) Ligands (yellow) bind to a macromolecule (green) with $N_B=2$ binding sites.  In the  $N=4$ state kinetic model, only the binding rates (pink) are proportional to the ligand concentration $c$.
(b) Decomposition of a $c$-perturbation into $E$-perturbations (blue tones) and a multi-edge $B$-perturbation (red tones).}
\label{fig:model}
\end{figure}

A central biochemical motif is the cooperative binding of ligands to a larger macromolecule~\cite{HillCoop,PhillipsMWC}. 
A relevant theoretical and experimental question is how sensitively this system responds to changes in the ligand concentration $c$, and how that depends on the number of binding sites $N_B$ as well as other structural and thermodynamic characteristics~\cite{Mahdavi2023}.
In a kinetic model, like the one in Fig.~\ref{fig:model}, each state $i$ is identified by the collection of binding sites in the macromolecule occupied by a ligand, with the number of bound sites denoted by the ligand-occupation number $n_i$.
Binding rates $k_\alpha c$ are taken to be proportional to $c$, whereas unbinding rates ${\tilde k}_\alpha$ are concentration independent; beyond that the rates are  fixed in accordance with local detailed balance \eqref{eq:cycle_force}.
It is then common to study normalized observables, $0\le f_i\le 1$: the fraction of bound sites $f_i=n_i/N_B$ is one example~\cite{Hill}, or in the context of gene regulation $f_i$ could represent a (normalized) transcription rate~\cite{RazoMejia2018,Reimer2023,Kim2022} .

Before addressing nonequilibrium situations, we recall the equilibrium limits to cooperative binding.
Statistical physics \eqref{eq:FDT_ener} predicts that the sensitivity of an arbitrary normalized observable $f_i$ is given by the correlation with the ligand-occupation number $n_i$,
\begin{equation}\label{eq:bind_eq}
\begin{split}
\frac{\partial \langle f\rangle_c^{\rm eq}}{\partial \ln c} &=\left\llangle f,n\right\rrangle^{\rm eq}\\
&\le N_B\langle f\rangle_c^{\rm eq}(1-\langle f\rangle_c^{\rm eq})\le   N_B/4.
\end{split}
\end{equation}
Here, we have bounded the covariance using \eqref{eq:operational} after recognizing $n_i/N_B\le 1$.
This allows us to recover the classic prediction that the slope of the binding curve is limited by the number of binding sites $N_B$~\cite{Hill}.  

When the system is driven away from equilibrium, binding can be more sensitive.
With the predictions presented above, we are limited to the model depicted in Fig.~\ref{fig:model}(a) with $N_B=2$.
To incorporate nonequilbrium driving, we imagine that the binding is coupled to ATP hydrolysis so that the sole thermodynamic driving force can be related to the chemical potential difference between ATP and its products: $F_{C} = \Delta\mu/kT$.

\begin{figure*}
\includegraphics[width=\textwidth]{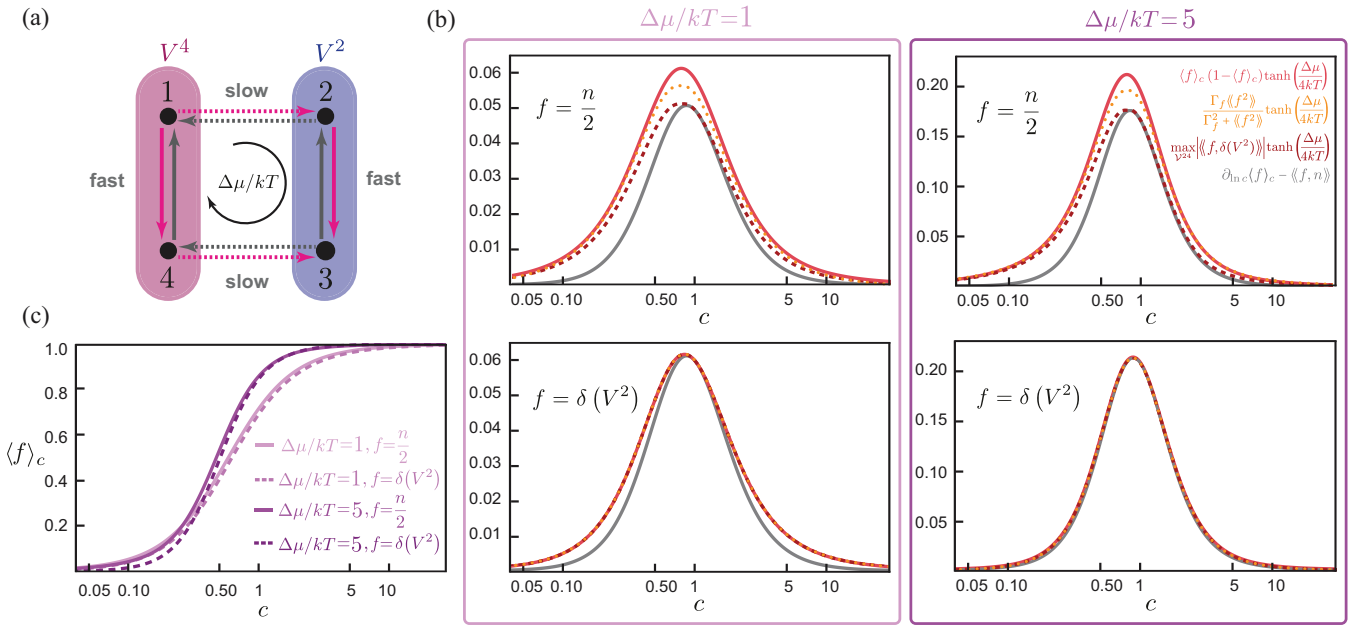}
\caption{Saturating the inequalities in receptor binding: (a) Representation of the optimal network corresponding to the topological consistent splitting is $V^2=\{1,4\}$, $V^2=\{2,3\}$. The rates internal to the islands (pink and blue ovals) are fast compared to the rates connecting the islands (dashed). (b) For the optimal network in (a), the deviation of the response from the equilibrium prediction (gray) is compared to the bounds imposed by \eqref{eq:f_bound} (dashed dark red), \eqref{eq:f_bound2} (orange) and \eqref{eq:f_bound3} (red) as a function of concentration $c$ for two observables $f=n/2$ (top) and $f=\delta\left(V^2\right)$ (bottom) at two values of thermodynamic driving, $\Delta\mu/kT=1$ (left) and  $\Delta\mu/kT=5$ (right). (c) $\langle f \rangle_c$ as a function of $c$ for both observables at the two values of thermodynamic driving. Parameter values are listed in Appendix \ref{sec:fig_values}.}
\label{fig:receptorBound}
\end{figure*}

Now, a logarithmic perturbation of $c$ can be divided into $E$-perturbations \eqref{eq:decomposition_E} and a multi-edge symmetric perturbation, as in Fig.~\ref{fig:model}(b).
 Thus, by combining \eqref{eq:E-perturbation} and \eqref{eq:bound_B_multi} we can bound the deviation of the sensitivity from the equilibrium-like prediction $\left\llangle f,n\right\rrangle$ (cf.\ \eqref{eq:bind_eq})
 \begin{equation}\label{eq:f_bound}
\left| \frac{\partial \langle f\rangle_c}{\partial \ln c} -  \left\llangle f,n\right\rrangle\right|\le \underset{{\mathcal V}^{24}}{\max}\Big|\left\llangle f,\delta(V^{2})\right\rrangle\Big|\tanh(\Delta\mu/4kT).
 \end{equation}
This can be simplified using the operational bounds \eqref{eq:operational3} and \eqref{eq:operational},
  \begin{align}\label{eq:f_bound2}
\left| \frac{\partial \langle f\rangle_c}{\partial \ln c} -  \left\llangle f,n\right\rrangle\right|&\le \frac{\Gamma_f \llangle f^2\rrangle}{\Gamma_f^2+\llangle f^2\rrangle} \tanh(\Delta\mu/4kT)\\
\label{eq:f_bound3}
&\le \langle f\rangle_c(1-\langle f\rangle_c)\tanh(\Delta\mu/4kT),
 \end{align}
 with $\Gamma_f = \min\{\langle f\rangle,1-\langle f\rangle, \sqrt{\llangle f^2\rrangle}\}$.
 We illustrate these inequalities in Fig.~\ref{fig:receptorBound} for the representative optimal topology in Fig.~\ref{fig:receptorBound}(a) where $V^2 = \{2,3\}$ and $V^4=\{1,4\}$.
Our bound singles out two normalized observables of potential interest: when the observable is proportional to the ligand-occupation number, $f_i= n_i/2,$ or when it is equal to the indicator function, $f_i=\delta_i(V^2)$.
For both observables, we plot in Fig.~\ref{fig:receptorBound}(b) the deviation of the response from the equilibirum-like expectation with the limits imposed by  \eqref{eq:f_bound} - \eqref{eq:f_bound3} as a function of the concentration $c$ for a lower-level of thermodynamic driving $\Delta\mu/kT=1$ and a higher-level of driving $\Delta\mu/kT=5$.
We observe that the covariance bound \eqref{eq:f_bound} is only saturated at one point.
Though, over the entire binding curve the bounds tend to be tighter with increasing thermodynamic driving, leading to an overall steeper binding curve as illustrated in Fig.~\ref{fig:receptorBound}(c).
Of the two observables, the bounds are tighter when the observable is equal to the indicator function $f_i=\delta_i(V^2)$, when further the observable's variance equals the operational limit, $\llangle \delta^2\rrangle=\langle \delta\rangle(1-\langle\delta\rangle)$.

\begin{figure*}
\includegraphics[width=\textwidth]{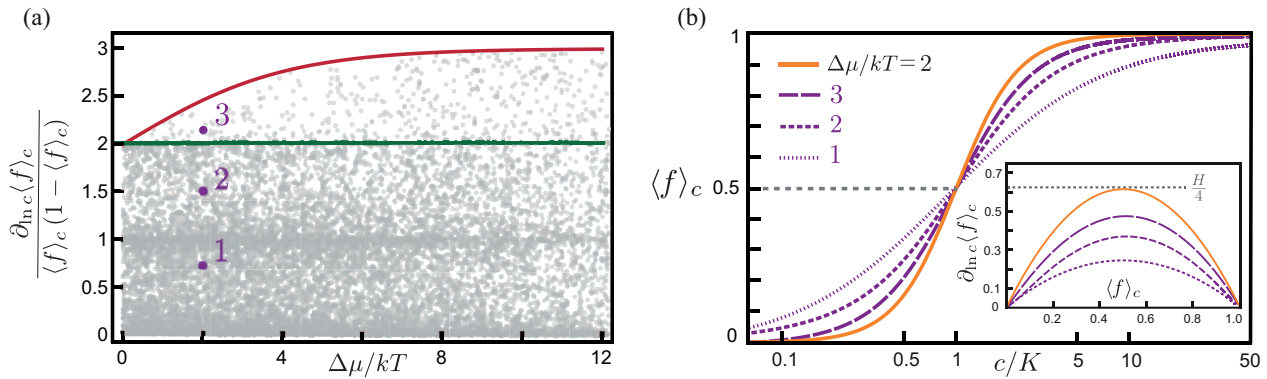}
\caption{Nonequilibrium response in the receptor binding model: (a) Normalized response for $15000$ random rate matrices with $c=1$, logarithmic asymmetric $\ln(k_\alpha/{\tilde k}_\alpha)$ and symmetric $\ln(k_\alpha{\tilde k}_\alpha)$ parts sampled uniformly on $[-7,7]$, 
and $f_1=0$, $f_3=1$, with $f_2=f_4$ uniform in $[0.4,0.6]$. All samples fall below the predicted bound~\eqref{eq:f_op_3}, red. Green line is the maximum equilibrium response $2$. (b) $\langle f \rangle_c$ as a function of $c/K$ for three distinct rate matrices [purple points (a)] with $\Delta\mu/kT = 2$, which need not be Hill functions. Orange curve corresponds to the optimal Hill function~\eqref{eq:f_op_hill}. Inset: for any $\langle f \rangle _c$ sensitivity is bounded by the steepness of the Hill function, with a maximum of $H/4$. For details refer to Appendix \ref{sec:fig_values}.}
\label{fig:hill_curves}
\end{figure*}

For observables that monotonically increase with concentration, \eqref{eq:f_bound} bounds the sensitivity
\begin{align}\label{eq:f_op_1}
\frac{\partial \langle f\rangle_c}{\partial \ln c} &\le \left\llangle f,n\right\rrangle +\underset{{\mathcal V}^{24}}{\max}\Big|\left\llangle f,\delta(V^{2})\right\rrangle\Big|\tanh(\Delta\mu/4kT)\\
\label{eq:f_op_2}
&\le \frac{\Gamma_f \llangle f^2\rrangle}{\Gamma_f^2+\llangle f^2\rrangle}\left(2+\tanh(\Delta\mu/4kT\right))\\
\label{eq:f_op_3}
&\le \langle f\rangle_c(1-\langle f\rangle_c)\left(2+\tanh(\Delta\mu/4kT\right)).
\end{align}
While these bounds saturate only at a single point along the binding curve, it is still worthwhile to investigate what concentration dependence would $\langle f\rangle_c$ require in order to have the maximum steepness, saturating \eqref{eq:f_op_2}, along the entire binding curve.
Assuming equality in \eqref{eq:f_op_2} leads to a differential equation for the optimal binding curve whose solution is the Hill function~\cite{Hill}
\begin{equation}\label{eq:f_op_hill}
\langle f\rangle^{\rm opt}_c  = \frac{c^H}{K^H+c^H}, \quad H =2+\tanh(\Delta\mu/4kT),
\end{equation}
where the Hill coefficient $H$ is enhanced beyond the structural equilibrium limit ($N_B=2$) by the chemical driving.
The arbitrary constant $K$ fixes the location of the curve via $\langle f\rangle^{\rm opt}_K=1/2$.
In Fig.~\ref{fig:hill_curves}(a), we verify the prediction in \eqref{eq:f_op_2}.
The optimality of the Hill function \eqref{eq:f_op_hill} is illustrated in Fig.~\ref{fig:hill_curves}(b).
Notably this curve does not bound any other binding curve, but for any given value of $\langle f\rangle$, it is the steepest.

\section{Conclusion} 
We have demonstrated that the maximum response over all rates holding the steady-state distribution fixed is given by the maximal fluctuations of the observable across the system's topologically-consistent splittings.
For symmetric edge perturbations this is further constrained by the degree of nonequilibrium driving as measured by the thermodynamic force.
These trade-offs quantify the role network topology plays in shaping the connection between response and fluctuations away from equilibrium.  

We can also view our predictions through the lens of the FDT.
In the FDT, the relevant metric of fluctuations is between the observable and a perturbation-dependent, conjugate observable, e.g., $U_i$ in \eqref{eq:FDT_ener}.
In the predictions derived here, the perturbation-dependent quantity that appears in the covariance is the indicator function for a topologically-consistent splitting.
In this way, the topologically-consistent splittings act as a kind of optimal conjugate observable, whose fluctuations help us organize our observations about response.

Our predictions suggest that the deep connection between response, fluctuations, and network structure can be fruitfully quantified in some generality.
One next step in this program is to go beyond single edge perturbation and incorporate correlated, multi-edge perturbations, which is necessary to apply our results to binding models with more than two sites~\cite{Estrada2016,Tran2018,Owen2023}.
Another direction is to move beyond the state variables we have analyzed to include current observables, which are themselves functions of the rates, like the inequality between mobility and diffusion derived in~\cite{Dechant2020,Baiesi2011}.
Finally, extending these predictions to time-dependent response could help in rationalizing observations about transcription in Eukaryotes where growing evidence is suggesting that timing is an important factor in regulation~\cite{Reimer2023, Fernandes2021}.  

\acknowledgements
This material is based upon work supported in part by the National Science Foundation under Grant No.\ 2142466 and by the Alfred P. Sloan Foundation under grant G-2022-19440.

\appendix
\begin{widetext}

\section{Graphical solutions for discrete markov processes}
\label{sec:Methods}

The Matrix-Tree Theorem (MTT) as well as its generalizations, the All Minors Matrix-Tree Theorem~\cite{Chaiken1982} or the Matrix-Forest Theorem~\cite{Chebotarev2002}, offer powerful graphical methods for organizing solutions of homogenous and inhomogeneous linear equations that arise in the context of discrete Markov processes.
These solutions are built by associating to a discrete Markov process with transition rate matrix $W=\{W_{ij}\}$, a weighted transition graph $G$ with vertices $\{i\}$ and \emph{directed} edges $\{e_{ij}\}$ weighted by $W_{ij}$.
Note that in the Appendices it will prove convenient to largely use directed edges, but label undirected edges as $\bar{e}_{ij}=\{e_{ij},e_{ji}\}$.
Then certain collections of subgraphs of $G$ will provide the desired solutions.

To this end, for any a subgraph $G'\subseteq G$, we will denote its vertex set as $V(G')$ and its edge set as $E(G')$.
We assign it a weight as the product of the weights of all its edges, $E(G') = \{i\to j, k\to l,\dots\}$, as
\begin{equation}
w(G') = W_{ji}W_{lk}\cdots.
\end{equation}
If a subgraph has no edges we will define its weight to be one. 
If it does not exist, its weight is zero.
Furthermore, the weight of a set of subgraphs ${\mathcal G}=\{G'_1,G'_2\dots\}$, is given by the sum over the weights of each subgraph, $w({\mathcal G}) = w(G_1)+w(G_2)+\cdots$.

Graphical solutions are then built out of spanning $n$-forests, which are subgraphs of $G$ composed of a collection of $n$ disjoint individually-connected components with no cycles such that each vertex is in exactly one component.  
In each component, we can choose a vertex, $r_1,\dots, r_n$, called the root, and orient all edges in each component along the unique path towards that root.
The resultant subgraph, we call a rooted spanning $n$-forest and will denote it as $f_{r_1}\sqcup f_{r_2}\sqcup\cdots f_{r_n}$.
When there is only one component, we will call them trees as opposed to $1$-forests and denote them as $T_r$.
To illustrate these concepts throughout, we will use the house graph (Fig.~\ref{fig:graph}), whose spanning trees and 2-forests are depicted in 
Fig.~\ref{fig:trees}.
\begin{figure}
    \centering
    \includegraphics[width=15.2cm]{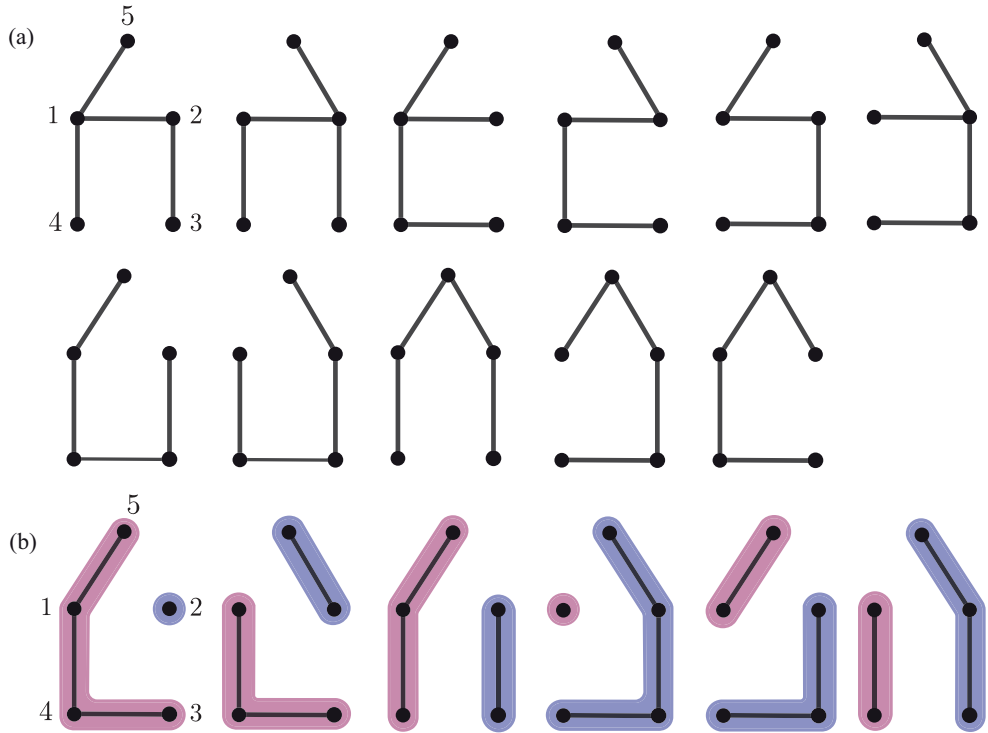}
    \caption{(a) All spanning trees, $\mathcal{T}$, for the house graph, Fig.~\ref{fig:graph}. (b) All spanning $2$-forests, $\mathcal{F}^{12}$, for the house graph. Note that there exist forest components that are composed of only one vertex. The representation of the topologically-consistent splitting of the vertices, $\mathcal{V}^{12}$, is also reproduced here. The set $V^1$, highlighted in pink, contains vertex $1$, and $V^2$, highlighted in blue, contains vertex $2$.} 
    \label{fig:trees}
\end{figure}

\subsection{Homogenous linear equations}
The MTT states that the steady-state solution of the master equation $\sum_j W_{ij}\pi_j = 0$ can be written as a sum over all spanning trees ${\mathcal T}$, 
\begin{equation}\label{eq:MTT}
    \pi_i = \frac{1}{\mathcal{N}}\sum_{\mathcal{T}}w\left(T_i\right),
\end{equation}
with normalization constant $\mathcal{N}= \sum_{\mathcal{T},k}w\left(T_k\right)=w({\mathcal T})$.
This is illustrated in Fig.~\ref{fig:MTT} for the house graph.

\begin{figure}
    \centering
    \includegraphics[width=15.2cm]{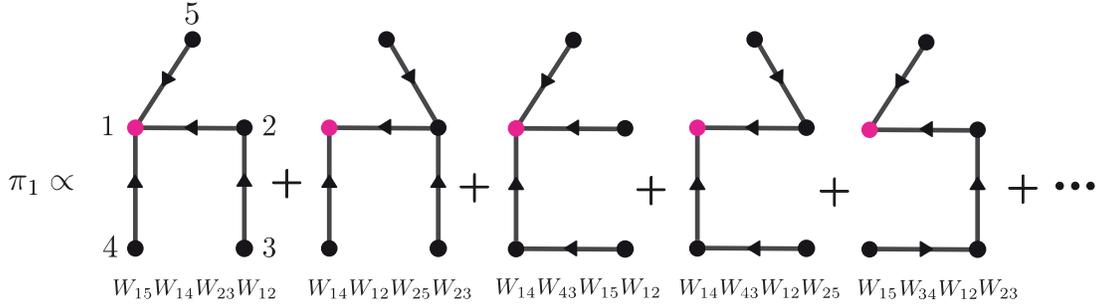}
    \caption{Graphical representation of the Matrix-Tree Theorem~\eqref{eq:MTT} for the house graph. The steady-state probability of any state $i$, in this case $i=1$, is given by a normalized sum of all the weights of spanning trees rooted at vertex $i$ (highlighted in pink).}
    \label{fig:MTT}
\end{figure}

Equation \eqref{eq:MTT} also leads to a graphical representation of the steady-state currents ${\bar J}_{mn} = W_{mn}\pi_n-W_{nm}\pi_m$~\cite{Hill}.
This representation is obtained by substituting \eqref{eq:MTT} into the definition of ${\bar J}_{mn}$.\
As explained in Ref.~\cite{Hill}, most terms cancel except for those that correspond to specific subgraphs of $G$, which we call (spanning) cycle graphs $C^{mn}$, formed by a single cycle, oriented along the edge $n\to m$, with every other vertex linked by a unique path oriented towards the cycle: they are formed by adding the edge $e_{mn}$ to a tree $T_n$ that did not have the edge $e_{nm}$.
When a cycle is oriented in the opposite direction we will at times write $\tilde{C}^{mn}=C^{nm}$.
The set of all cycle graphs we denote ${\mathcal C}^{mn}$.
Then the steady-state current is the difference of oriented cycle graphs 
\begin{equation}
J_{mn} = \frac{1}{{\mathcal N}} \sum_{C^{mn}} w(C^{mn})-w(\tilde{C}^{mn}) = \frac{1}{\mathcal N} \Big[w({\mathcal C}^{mn})- w({\mathcal C}^{nm})\Big].
\end{equation}

This formulation also offers a deep connection with the cycle forces.
Noting that the only difference between a cycle graph $C^{mn}$ and its reverse $\tilde{C}^{mn}$ are the weights along the cycle, allows us to identify the cycle forces as
\begin{equation}\label{eq:force}
F(C^{mn}) = \ln \frac{w(C^{mn})}{w(\tilde{C}^{mn})},
\end{equation}
because all the edge weights not in the cycle cancel. 
Substituting into the steady-state current,
\begin{equation}
J_{mn} = \frac{1}{{\mathcal N}} \sum_{C^{mn}} w(C^{mn})\left(1-e^{-F(C^{mn})}\right),
\end{equation}
leads to the conclusion that nonzero currents require nonzero cycle forces, confirming our expectation that nonequilibrium flows require driving~\cite{Hill}.

\subsection{Inhomogenous linear equations}
The other class of linear equations we will confront are of the form
\begin{equation}\label{eq:equation_forest}
\sum_{j=1}^N W_{ij} r_j = \delta_{im}-\delta_{in},
\end{equation}
with the additional condition $\sum_j r_j = 0$. 
A graphical solution to these inhomogenous equations can be developed in terms of the $2$-forests $f^m\sqcup f^n\in {\mathcal F}^{mn}$, where component $f^m$ has vertex $m$ and component $f^n$ has vertex $n$,
\begin{equation}\label{eq:solution_forest}
    r_j=\frac{1}{\mathcal{N}}\sum_{\mathcal{F}^{mn},l}w\left(f_{l}^{m}\sqcup f_{j}^{n}\right)-w\left(f_{j}^{m}\sqcup f_{l}^{n}\right).
\end{equation}
Here, we follow the solution method presented in Ref.~\cite{Caplan1982}, based on Hill's graphical proof of the MTT~\cite{Hill}.  
A slight modification of the argument is needed, and for completeness we include it in the following.
This approach can be seen to complement more general statements based on algebraic arguments found in Refs.~\cite{Chaiken1982,Chebotarev2002}.

To see that \eqref{eq:solution_forest} solves \eqref{eq:equation_forest}, we note that \eqref{eq:equation_forest} can be divided into three 
different types of equations depending on the value of $i$:
\begin{align}
&\sum_{j\neq m}W_{mj}r_{j}-\sum_{j\neq m}W_{jm}r_{m}=1,\label{eq:type_1}\\
&\sum_{j\neq n}W_{nj}r_{j}-\sum_{j\neq n}W_{jn}r_{n}=-1,\label{eq:type_2}\\
&\sum_{j\neq h}W_{hj}r_{j}-\sum_{j\neq h}W_{jh}r_{h}=0, \label{eq:type_3} \qquad h\neq m,n.
\end{align}
where conservation of probability is used to expand the diagonal elements of the rate matrix, $W_{ii}=-\sum_{j\neq i}W_{ji}$.

We now check that \eqref{eq:solution_forest} is the solution by direct calculation.
First, substituting into \eqref{eq:type_1}, we arrive at the pair of sums
\begin{align}\label{eq:type_1a}
\sum_{j\neq m} W_{mj} r_j &= \frac{1}{{\mathcal N}}\sum_{\substack{{\mathcal F}^{mn},l \\ j\neq m}} W_{mj} \left[w(f_{l}^{m}\sqcup f_j^n)-w(f_{j}^{m}\sqcup f_{l}^{n})\right]\\
\label{eq:type_1b}
-\sum_{j\neq m} W_{jm} r_m &=  \frac{1}{{\mathcal N}}\sum_{\substack{{\mathcal F}^{mn},l \\ j\neq m}}W_{jm}w(f^m_m\sqcup f^n_l).
\end{align}
Now, each term in \eqref{eq:type_1a} and \eqref{eq:type_1b} corresponds to the weight of a subgraph formed from the addition of a single edge to a 2-forest.
We address each sum in turn.
The terms of the form $W_{mj}w(f_l^m\sqcup f_j^n)$ correspond to the weight of a subgraph constructed by adding $e_{mj}$ to the 2-forest $f_l^m\sqcup f_j^n$, which forms a rooted spanning tree such that along the unique sequence of undirected edges linking $n$ and $m$, the edge incident to $m$ is oriented into $m$.
We call the set of trees with this property ${\mathcal T}^{m \leftlsquigarrow n}$.   
All trees in the set are formed in this manner: the removal of the unique edge into $m$ leads to the  2-forest $f_l^m\sqcup f_j^n$.
Next, the terms in the second sum of \eqref{eq:type_1a} have the form $W_{mj}w(f_j^m\sqcup f_l^n)$.
Here, the addition of the edge $e_{mj}$ to the 2-forest $f_j^m\sqcup f_l^n$ closes a cycle in the $f^m$ component, leaving the $f^n$ component unaltered. 
This set of subgraphs we denote ${\mathcal C}^{m}{\mathcal F}^n$, where again the entire set is formed via this construction.
The last type of term $W_{jm}w(f^m_m\sqcup f^n_l)$ appearing in \eqref{eq:type_1b}, corresponds to the addition of $e_{jm}$ to $f_m^m\sqcup f_l^n$, leading to one of two possibilities.
The first possibility is that $j\in V(f^m)$ is in the same component as $m$.
In this case, the resulting subgraph is an element of ${\mathcal C}^{m}{\mathcal F}^n$, the sum in \eqref{eq:type_1b} spanning over the entire set.
The other possibility is that $j\notin V(f^m)$ is not in the same component as $m$, and $e_{jm}$ links the two components of the forest forming a rooted spanning tree with the property that in the unique sequence of undirected edges linking $n$ and $m$ the edge incident to $m$ is oriented out of $m$.
This set we call ${\mathcal T}^{m \rightsquigarrow n}$.
Putting everything together, we can write \eqref{eq:type_1} as
\begin{equation}
\sum_{j\neq m}W_{mj}r_{j}-\sum_{j\neq m}W_{jm}r_{m}=\frac{1}{\mathcal N}\left[w\left({\mathcal T}^{m \leftlsquigarrow n}\right)-w\left({\mathcal C}^{m}{\mathcal F}^n\right)+w\left({\mathcal T}^{m \rightsquigarrow n}\right)+w\left({\mathcal C}^{m}{\mathcal F}^n\right)\right] = \frac{w({\mathcal T})}{{\mathcal N}} = 1,
\end{equation}
where we recognized that ${\mathcal T}^{m \leftlsquigarrow n}\cup{\mathcal T}^{m \rightsquigarrow n}$, since in any tree, along the path linking $n$ and $m$ the edge incident to $m$ can only be oriented either into or out of $m$.
Verification that \eqref{eq:solution_forest} solves \eqref{eq:type_2} follows along similar lines, so we omit the argument.

Lastly, we have to check \eqref{eq:type_3}.
We proceed by analyzing the pair of sums
\begin{align}\label{eq:type_3a}
\sum_{j\neq h} W_{hj}r_j&=\frac{1}{{\mathcal N}}\sum_{\substack{{\mathcal F}^{mn},l \\ j\neq h}} W_{hj} \left[w(f_{l}^{m}\sqcup f_j^n)-w(f_{j}^{m}\sqcup f_{l}^{n})\right]\\
\label{eq:type_3b}
-\sum_{j\neq h} W_{jh}r_h &=-\frac{1}{{\mathcal N}}\sum_{\substack{{\mathcal F}^{mn},l \\ j\neq h}} W_{jh} \left[w(f_{l}^{m}\sqcup f_h^n)-w(f_h^{m}\sqcup f_{l}^{n})\right]
\end{align}
Again, sums on the right hand sides correspond to the weights of a set of subgraphs.
The three sets that emerge are
\begin{enumerate}
\item ${\mathcal T}^{n {\buildrel h\over \leftrightsquigarrow} m}$: spanning trees such that the sequence of undirected edges connecting $n$ and $m$ pass through $h$, obtained by adding an edge incident to $h$ that joins the two components of a 2-forest.
\item ${\mathcal C}^{m,h}{\mathcal F}^n$: spanning subgraphs composed of a cycle graph component containing the vertex $m$ and a cycle through $h$, plus a tree component containing the vertex $n$.  They are formed by adding an edge incident to $h$ to a 2-forest  $f^m\sqcup f^n$, where the edge links two vertices in the $f^m$ component.
\item ${\mathcal F}^m{\mathcal C}^{n,h}$: spanning subgraphs composed of a cycle graph component containing the vertex $n$ and a cycle through $h$, plus a tree component containing the vertex $m$.  They are formed by adding an edge incident to $h$ to a 2-forest  $f^m\sqcup f^n$, where the edge links two vertices in the $f^n$ component.
\end{enumerate}
With this notation we separately combine the positive and negative terms in the sums in \eqref{eq:type_3a} and \eqref{eq:type_3b}, so that they can be equated to the weights of these sets as
\begin{align}\label{eq:type_3c}
&\sum_{\substack{{\mathcal F}^{mn},l \\ j\neq h}} W_{hj} w(f_{l}^{m}\sqcup f_j^n)+W_{jh}w(f_h^{m}\sqcup f_{l}^{n}) = w\left({\mathcal F}^m{\mathcal C}^{n,h}\right)+w\left({\mathcal C}^{m,h}{\mathcal F}^n\right)+w\Big({\mathcal T}^{n {\buildrel h\over \leftrightsquigarrow} m}\Big),\\
\label{eq:type_3d}
-&\sum_{\substack{{\mathcal F}^{mn},l \\ j\neq h}} W_{hj} w(f_{j}^{m}\sqcup f_l^n)+W_{jh}w(f_l^{m}\sqcup f_h^{n}) =-w\left({\mathcal C}^{m,h}{\mathcal F}^n\right)- w\left({\mathcal F}^m{\mathcal C}^{n,h}\right)-w\Big({\mathcal T}^{n {\buildrel h\over \leftrightsquigarrow} m}\Big).
\end{align}
It is easy to see that both sums generate the entire sets ${\mathcal F}^m{\mathcal C}^{n,h}$ and ${\mathcal C}^{m,h}{\mathcal F}^n$.
Furthermore, every element of ${\mathcal T}^{n {\buildrel h\over \leftrightsquigarrow} m}$ is indeed contained in the sums on the left hand sides.
To see this, let us follow the sequence of undirected edges from $m$ to $n$ in order.  
If we remove the unique edge immediately following vertex $h$, we obtain a 2-forest where $m$ and $h$ are in the same component, which corresponds to terms in one of the two sums in \eqref{eq:type_3c}, either $W_{hj} w(f_j^m\sqcup f_l^n)$ or $W_{jh} w(f_h^m\sqcup f_l^n)$ depending on the orientation of the edge.
Similarly, if we remove the edge immediately prior to the vertex $h$, we obtain a term in \eqref{eq:type_3d} where $n$ and $h$ share a component.
Clearly, the sum of \eqref{eq:type_3c} and $\eqref{eq:type_3d}$ is zero completing the argument.

\section{Symmetric edge perturbation}\label{sec:symmetric}
In this section, we prove the thermodynamic bound in~\eqref{eq:bound_B}.
The plan is to first utilize the graphical solutions to linear equations described above to arrange the expression for the observable response into a convenient form.
This form will allow us to state our problem of bounding the response as a linear optimization problem, whose optima provide the desired limits.

\subsection{Response as a linear optimization problem}
To begin, we observe that the response of the average $\langle Q\rangle = \sum_i Q_i\pi_i$ of an observable,
\begin{equation}\label{eq:Qderivative}
  \frac{\partial\langle Q\rangle}{\partial B_{mn}} = \sum_{i=1}^N Q_i  \frac{\partial\pi_{i}}{\partial B_{mn}},
\end{equation}
 is determined by how the steady-state distribution responds.
By differentiating the master equation~\eqref{eq:master}, we find that these derivatives can be obtained as the solution of the inhomogeneous linear equations
\begin{equation}
\sum_{j=1}^N W_{ij}  \frac{\partial\pi_{j}}{\partial B_{mn}} = {\bar J}_{mn}(\delta_{im}-\delta_{in}).
\end{equation}
These equations have the form previously introduced in \eqref{eq:equation_forest}, and thus the solution can be compactly organized in terms of 2-forests \eqref{eq:solution_forest} as
\begin{equation}\label{eq:solution_B_main}
    \frac{\partial\pi_{i}}{\partial B_{mn}}=\frac{\bar{J}_{mn}}{\mathcal{N}}\sum_{\mathcal{F}^{mn},j}w\left(f_{j}^{m}\sqcup f_{i}^{n}\right)-w\left(f_{i}^{m}\sqcup f_{j}^{n}\right),
\end{equation}
\noindent which is illustrated in Fig.~\ref{fig:response_forests}.

\begin{figure}
    \centering
    \includegraphics[width=15.2cm]{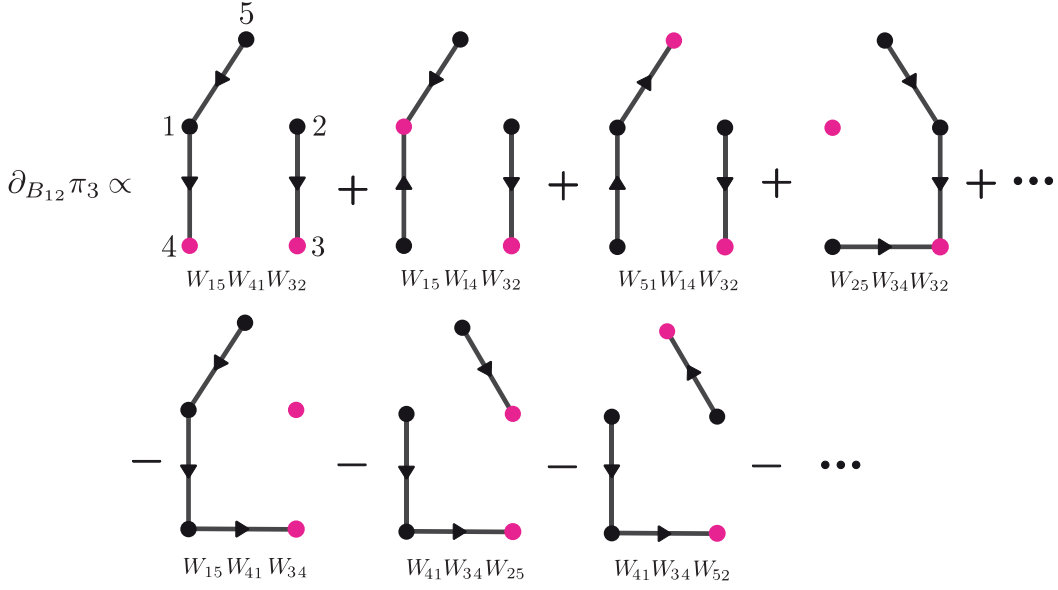}
    \caption{Graphical representation of the solution to the class of linear equations~\eqref{eq:solution_forest}, which provides a convenient expression for the symmetric edge perturbation~\eqref{eq:solution_B_main} of the steady-state probability distribution. We again consider the house graph with a symmetric edge perturbation on edge $\bar e_{12}$. The response of $\pi_3$, is composed of two terms, a positive one, which is a sum of weights of all rooted $2$-forests $f^1_l \sqcup f^2_3$ (both roots are highlighted in pink), and a negative one, which is a sum of weights on all rooted $2$-forests $f^1_3\sqcup f^2_l$.}
    \label{fig:response_forests}
\end{figure} 

Inserting this expression into \eqref{eq:Qderivative}, rearranging the sum, then multiplying and dividing by two factors---$\pi_i\pi_j$ and $\left(\sqrt{w({\mathcal C}^{mn})}+\sqrt{w({\mathcal C}^{nm})}\right)^2$---we arrive at the sought after form 
\begin{equation}\label{eq:expression_linOpt}
   \left| \frac{\partial\langle Q\rangle}{\partial B_{mn}}\right|={\mathcal M} \bigg| \sum_{i,j}\left(Q_j-Q_i\right)\pi_i\pi_j {\mathcal P}_{ij}\bigg|,
\end{equation}
where we have separated out an overall magnitude
\begin{equation}\label{eq:M}
{\mathcal M} = \frac{ \left|w({\mathcal C}^{mn})-w({\mathcal C}^{nm})\right|}{\left(\sqrt{w({\mathcal C}^{mn})}+\sqrt{w({\mathcal C}^{nm})}\right)^2},
\end{equation}
and a collection of structural coefficients
\begin{equation}\label{eq:struc_coeff}
{\mathcal P}_{ij} = \left(\sqrt{w({\mathcal C}^{mn})}+\sqrt{w({\mathcal C}^{nm})}\right)^2  P_{ij},\qquad{\rm where}\qquad  P_{ij} = \frac{\sum_{\mathcal{F}^{mn}}w(f_{i}^{m}\sqcup f_{j}^{n})}{\sum_{\mathcal T}w(T_i)\sum_{\mathcal T}w(T_j)}.
\end{equation}
are forest-tree ratios.  
Note that the subscripts align with the forest notation, so that the left subscript $i$ is in the $f^m$ component and right subscript $j$ is in the $f^n$ component. 

Equation~\eqref{eq:expression_linOpt} allows us to divide the problem in two.
In Appendix~\ref{sec:mag}, we show that the overall magnitude is bounded by the thermodynamic driving force,
\begin{equation}
{\mathcal M} \le \tanh({\mathcal F}_{\rm max}/4).
\end{equation}
where ${\mathcal F}_{\rm max} = \max_{C^{mn}}|F(C^{mn})|$ is the maximum over all cycle forces through the perturbed edge.
Then in Appendix~\ref{sec:struc}, we use graph-theoretic arguments to deduce linear relationships among the structural coefficients, demonstrating that the structural coefficients ${\mathcal P}_{ij}$ are confined to a convex polytope.
As result, the fundamental theorem of linear programing~\cite{Luenberger} implies that the vertices of this polytope, which turn out to be topologically-consistent splittings, are potential optima, and  
\begin{equation}\label{eq:struc_bound}
\underset{{\mathcal P}_{ij}}{\max} \bigg| \sum_{i,j}\left(Q_j-Q_i\right)\pi_i\pi_j {\mathcal P}_{ij}\bigg| \le \underset{\mathcal{V}^{mn}}{\max} \Big|\llangle Q,\delta(V^{m})\rrangle\Big|.
\end{equation}
Together these inequalities imply our main result~\eqref{eq:bound_B}.

\subsection{Bounding the response magnitude}\label{sec:mag}

Factoring the numerator of the response magnitude \eqref{eq:M}, we find
\begin{equation}
{\mathcal M} = \left|\frac{\sqrt{w({\mathcal C}^{mn})}-\sqrt{w({\mathcal C}^{nm})}}{\sqrt{w({\mathcal C}^{mn})}+\sqrt{w({\mathcal C}^{nm})}}\right| = \tanh\left(\frac{1}{4}\left|\ln\frac{w({\mathcal C}^{mn})}{w({\mathcal C}^{nm})}\right| \right).
\end{equation}
Then an application of the log-sum inequality~\cite{Cover} to the sums over cycle graphs and the definition of cycle force \eqref{eq:force}, leads to the sequence of bounds
\begin{equation}\label{eq:responseMagnitudeBound}
{\mathcal M}\le \tanh\left(\frac{1}{4\, w({\mathcal C}^{mn})}\sum_{C^{mn}} w(C^{mn})\left| F(C^{mn})\right|\right)\le \tanh({\mathcal F}_{\rm max}/4).
\end{equation}

\subsection{Bounding the structural coefficients}\label{sec:struc}

The structural coefficients \eqref{eq:struc_coeff} are formed from spanning forests and therefore are not independent.
In this section, we will first demonstrate that the forest-tree ratios $P_{ij}$ (and therefore the structural coefficients ${\mathcal P}_{ij}$) are constrained by a collection of linear equalities and inequalities, and thus are confined to a convex polytope.  
This will allow us to apply the  machinery from optimization theory to deduce our bounds on response.

First, we note that due to their definition, the forest-tree ratios are nonnegative, $P_{ij}\ge 0$.
A number of them are also trivially zero,
\begin{align}
    & P_{im} = 0  \:\:\:\: i\neq m, \label{eq:exc_P_im}\\
    & P_{ni} = 0  \:\:\:\:\: i\neq n, \label{eq:exc_P_ni} \\
    & P_{ii} = 0, \label{eq:exc_P_ii}
\end{align}
since a vertex can only be in one component, and the 2-forests $f_j^m\sqcup f^n_m$ and $f_n^m\sqcup f^n_i$ with $m$ or $n$ in the opposite component are not possible.
In the following we will show that the forest-tree ratios are further related by
\begin{align}\label{eq:struc_const_1}
&P_{ij}\le P_{mj}, P_{in} \\
\label{eq:struc_const_2}
&P_{mi}+P_{ij} = P_{mj}+P_{ji} \\
\label{eq:struc_const_3}
&P_{in}+P_{ji} = P_{jn}+P_{ij}
\end{align}
which naturally extend to the structural coefficients ${\mathcal P}_{ij}$ that determine the response. 
Furthermore, we will demonstrate that the structural coefficient 
\begin{equation}\label{eq:struc_const_4}
{\mathcal P}_{mn}\le 1.
\end{equation}
is bounded, which also constrains all other structural coefficients ${\mathcal P}_{ij}\le 1$ \eqref{eq:struc_const_1}.\\

Our main tool for  deducing the above relationships we call the \emph{Root-Swap} map.
It is an invertible function that maps a rooted spanning tree and a rooted spanning $2$-forest $(T_i,f^m_k\sqcup f^n_l)$ to another such pair keeping the total weight fixed, but interchanging the roots of the tree and the forest:
\begin{equation}
w(T_i)w(f^m_k\sqcup f^n_l) \xrightarrow[]{\emph{Root-Swap}} \Big\{ w(T'_k)w(f'^m_i\sqcup f'^n_l)\quad {\rm or}\quad  w(T'_l)w(f'^m_k\sqcup f'^n_i)\Big\}.
\end{equation}
Notably, which root of the forest ends up getting swapped depends on the input tree $T_i$.
Though no matter the case, the output of the map is unique.

To construct this map,  we need to introduce additional definitions:
\begin{quote}
\textit{Source of directed edge}: denoted $s\left(f\right)$ --- the starting vertex of a directed edge $f$; 

\textit{Target of directed edge}: denoted $t\left(e\right)$ --- the ending vertex of a directed edge $e$; 

\textit{Doubly-rooted spanning $2$-forest}: denoted $f_{k}^{m}\sqcup f_{lp,q}^{n}$ --- a subgraph of $G$ which is a spanning $2$-forest formed as follows. The first component $f_{k}^{m}$ contains vertex $m$ and is rooted at $k$. The second component $f_{lp,q}^{n}$ contains vertex $n$ and is doubly-routed with roots $l$ and $p$ split by a branch point $q$, i.e., every edge is directed as in $f_{l}^{n}$ and $f_{p}^{n}$ when those directions coincide, and otherwise directed toward $l$ if between $q$ and $l$ and toward $p$ if between $q$ and $p$. Note that $f_{k}^{m}\sqcup f_{l}^{n}$, coincides with any doubly-rooted $2$-forest of the type $f_{k}^{m}\sqcup f_{pl,p}^{n}$ or $f_{k}^{m}\sqcup f_{lp,p}^{n}$ for any $p$. 
\end{quote}
\noindent The \emph{Root-Swap} map is then built from repeated applications of the \emph{Edge-Swap} operation:

\begin{quote}
\textit{Edge-Swap}: Input $\left(T_{q},f_{k}^{m}\sqcup {f}_{lp,q}^{n}\right)$. Remove from $f_{lp,q}^{n}$ the unique edge $e$ pointing out of the branch point $q$ towards $p$, the vertex in the middle position. This splits ${f}_{lp,q}^{n}$ into two disjoint components: $D$,
which contains vertex $n$, and ${\bar D}$, which does not.
Insert $e$ into $T_{q}$, thereby creating a cycle $C$ oriented from $q=s(e)$ to $t\left(e\right)$. Note that this cycle may be formed by only two edges, $e$ and another edge in the opposite orientation.
Starting from $t(e)$, march along the links of this cycle $C$ following its orientation, until you find the first edge $f$ that reconnects ${\bar D}$ back to $D$ or to the other component $f^m_k$.
Remove $f$ to form a new rooted tree $T'_{s(f)}$, and insert $f$ back into the pieces of the $2$-forest to form either $f'^m_{kp,s(f)}\sqcup f'^n_l$, $f'^m_k \sqcup f'^n_{lp,s(f)}$, or $f'^m_{lk,s(f)}\sqcup f'^n_p$. Output the result: $\left(T'_{s\left(f\right)},{f'}_{kp,s\left(f\right)}^{m}\sqcup{f'}_{l}^{n}\right)$, $\left(T'_{s\left(f\right)},{f'}_{k}^{m}\sqcup{f'}_{lp,s\left(f\right)}^{n}\right)$ or $\left(T'_{s\left(f\right)},{f'}_{lk,s\left(f\right)}^{m}\sqcup{f'}_{p}^{n}\right)$. 
\end{quote}

\noindent With these tools in hand, we implement the \emph{Root-Swap} map on the pair $(T_i,f^m_k\sqcup f^n_l) $ as follows (where we assume without loss of generality that $i\in V\left(f^{n}\right)$ initially). An illustration is presented in Fig.~\ref{fig:root_swap}:
\begin{quote}
\emph{Root-Swap}$(T_i,f^m_k\sqcup f^n_l)$:\\
(1)\ Identify the input with the doubly-rooted forest as $\left(T_i,f_{k}^{m}\sqcup f_{il,i}^{n}\right)$. \\
(2)\ Repeatedly apply \emph{Edge-Swap} until there is no edge pointing out of the branch point in the appropriate direction, which occurs when the  
branch point becomes one of the original roots with pair $(T'_k, {f'}^m_{ik,k}\sqcup {f'}^n_{l})$ or $(T'_l, {f'}^m_{k}\sqcup {f'}^n_{il,l})$.\\
(3)\ Output the result, either $\left(T'_k,{f'}_{i}^{m}\sqcup {f'}_{l}^{n}\right)$ or $\left(T'_l,{f'}_{k}^{m}\sqcup {f'}_{i}^{n}\right)$.
\end{quote}

The above algorithm always terminates.
When the initial edge $e$ is removed during the first application of \emph{Edge-Swap}, the component $D$ containing $n$ may also include the original root $l$, or the original root may be in ${\bar D}$ with $D$ instead rooted at $i$.
In the first instance we initially have tree-components $D_l$ and ${\bar D}_i$, every subsequent application of \emph{Edge-Swap} will grow the component ${\bar D}_i$ by marching the branch point closer to $l$ or $k$, until the branch point merges with one of the original roots, and the algorithm terminates.
In the second instance, after removal of $e$ we have components $D_i$ and ${\bar D}_l$, with each application of \emph{Edge-Swap} shrinking ${\bar D}_l$ until the branch point merges with $l$, terminating the algorithm. 
This second instance is what is illustrated in Fig.~\ref{fig:root_swap}. 

\emph{Root-Swap} is also invertible. Each step of \emph{Edge-Swap} can be reversed, as we can always find the swapped edge by following the cycle $C$ along its reverse orientation. Moreover, the starting point can be identified since the root of the tree serves as the first branch point, which we see by looking at the terminal configurations, $(T'_k, {f'}^m_{ik,k}\sqcup {f'}^n_{l})$ or $(T'_l, {f'}^m_{k}\sqcup {f'}^n_{il,l})$. We would begin by removing the unique edge pointing into $k$ or $l$.

Finally the weights are conserved, as edges are merely swapped; no edges are created, destroyed or reoriented.
\newline
\begin{figure}
     \centering
     \includegraphics[width=14.2cm]{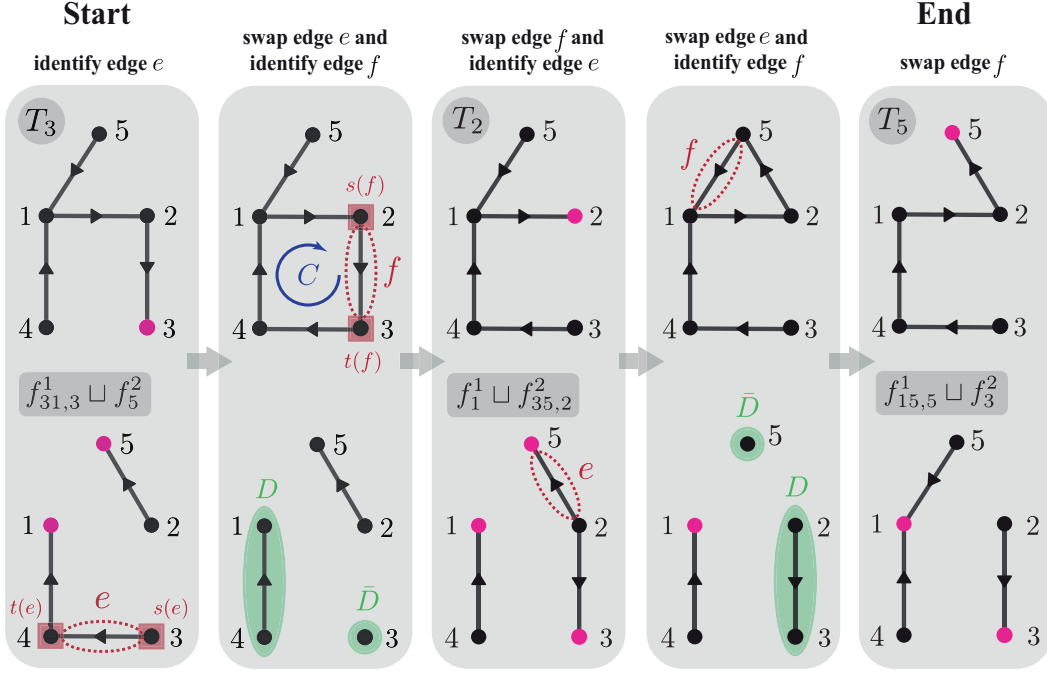}
     \caption{Example application of the \emph{Root-Swap} map taking the pair $(T_3,f^1_1\sqcup f^2_5)$ as input and outputting the pair $(T'_5,f'^1_1\sqcup f'^2_3)$.  		Each panel contains one step of the \emph{Edge-Swap} operation.}
     \label{fig:root_swap}
\end{figure}

With \emph{Root-Swap} in hand we now utilize it to derive \eqref{eq:struc_const_1}-\eqref{eq:struc_const_4}:\\

\noindent 1. \emph{Derivation of $P_{ij}\le P_{mj}, P_{in}$ \eqref{eq:struc_const_1}}:

The derivation proceeds by using the \emph{Root-Swap} map to change the subscripts in the forest-tree weights. 
With this in mind, we begin by manipulating the definition of $P_{mj}$ \eqref{eq:struc_coeff} by multiplying and dividing by the weight all spanning trees rooted at $i$, and then apply the \emph{Root-Swap} map:
\begin{align}\label{eq:struc_const_1a}
P_{mj} &= \frac{1}{\sum_{\mathcal T}w(T_m)\sum_{\mathcal T}w(T_i)\sum_{\mathcal T}w(T_j)}\left[\sum_{\mathcal T}\sum_{\mathcal{F}^{mn}}w(T_i)w(f_{m}^{m}\sqcup f_{j}^{n})\right] \\
\label{eq:struc_const_1b}
& = \frac{1}{\sum_{\mathcal T}w(T_m)\sum_{\mathcal T}w(T_i)\sum_{\mathcal T}w(T_j)}\left[\sum_{\{T_m, f^m_i\sqcup f^n_j\}\in {\mathcal RS}}w(T_m)w(f_i^{m}\sqcup f_j^{n})+\sum_{\{T_j, f^m_m\sqcup f^n_i\}\in {\mathcal RS}}w(T_j)w(f_{m}^{m}\sqcup f_{i}^{n})\right]
\end{align}
Now importantly, application of the \emph{Root-Swap} map does not necessarily generate every weighted product of trees and forests, but in general only a subset of ${\mathcal T}\times  {\mathcal F}^{mn}$.
This is notated in \eqref{eq:struc_const_1b} by confining the sum to the set  ${\mathcal RS}$ of tree-forest pairs generated by application of \emph{Root-Swap} to \eqref{eq:struc_const_1a}.
We now claim that in fact all pairs in the first term, which are of the form $w(T_m)w(f^m_i\sqcup f^n_j)$, are generated under \emph{Root-Swap} and that the sum actually extends over all tree-forest pairs.
This follows by imagining there is a tree-forest pair $w(T_m)w(f^m_i\sqcup f^n_j)$ not generated by \emph{Root-Swap} (not in ${\mathcal RS}$).  
We can then apply the inverse \emph{Root-Swap} map, which can only generate a pair of the form $w(T_i)w(f^m_m\sqcup f^n_j)$ (as $m$ cannot be in the $f^n$ component). 
Due to the uniqueness of the \emph{Root-Swap} map, this term had to be present in the original sum in \eqref{eq:struc_const_1a}, and therefore there are no pairs $w(T_m)w(f^m_i\sqcup f^n_j)$ not in ${\mathcal RS}$.
Thus, we have 
\begin{align}
P_{mj} & = \frac{1}{\sum_{\mathcal T}w(T_m)\sum_{\mathcal T}w(T_i)\sum_{\mathcal T}w(T_j)}\left[\sum_{{\mathcal T},{\mathcal F}^{mn}}w(T_m)w(f_i^{m}\sqcup f_j^{n})+\sum_{\{T_j, f^m_m\sqcup f^n_i\}\in {\mathcal RS}}w(T_j)w(f_{m}^{m}\sqcup f_{ji}^{n})\right] \\
&\ge P_{ij},
\end{align}
where we have identified $P_{ij}$ \eqref{eq:struc_coeff} and noted the remaining terms are positive.
A similar argument leads to the conclusion that $P_{in}\ge P_{ij}$ as well.\\

\noindent 2. \emph{Derivation of $P_{mi}+P_{ij} = P_{mj}+P_{ji}$ \eqref{eq:struc_const_2}} \& $P_{in}+P_{ji} = P_{jn}+P_{ij}$ \eqref{eq:struc_const_3}:

We rearrange the sum $P_{mi}+P_{ij}$ as
\begin{equation}\label{eq:struc_const_2a}
P_{mi}+P_{ij} =\frac{1}{\sum_{\mathcal T}w(T_m)\sum_{\mathcal T}w(T_i)\sum_{\mathcal T}w(T_j)} \left[\sum_{{\mathcal T},\mathcal{F}^{mn}}w(T_j)w(f_{m}^{m}\sqcup f_{i}^{n})+ \sum_{{\mathcal T},\mathcal{F}^{mn}}w(T_m)w(f_i^{m}\sqcup f_j^{n})\right],
\end{equation}
so that we can apply the \emph{Root-Swap} map to interchange the subscripts
\begin{equation}
P_{mi}+P_{ij} =\frac{1}{\sum_{\mathcal T}w(T_m)\sum_{\mathcal T}w(T_i)\sum_{\mathcal T}w(T_j)} \left[\sum_{\{T_m,f^m_j\sqcup f^n_i\}\in {\mathcal RS}'}w(T_m)w(f_{j}^{m}\sqcup f_{i}^{n})+\sum_{\{T_i,f^m_m\sqcup f^n_j\}\in {\mathcal RS}'} w(T_i)w(f_{m}^{m}\sqcup f_{j}^{n})\right],
\end{equation}
with ${\mathcal RS}'$ the image of the \emph{Root-Swap} map applied to \eqref{eq:struc_const_2a}.
Again, we claim that the sums actually extend over all tree-forest pairs, ${\mathcal T}\times{\mathcal F}^{mn}$.
Indeed, one can check, for example, if there were a pair $w(T_m)w(f_{j}^{m}\sqcup f_{i}^{n})\notin{\mathcal RS}'$, application of the inverse \emph{Root-Swap} map would generate a term in one of the sums in \eqref{eq:struc_const_2a}, leading to a contradiction.
Thus, we have 
\begin{align}
P_{mi}+P_{ij} &=\frac{1}{\sum_{\mathcal T}w(T_m)\sum_{\mathcal T}w(T_i)\sum_{\mathcal T}w(T_j)} \left[\sum_{{\mathcal T},\mathcal{F}^{mn}}w(T_m)w(f_{j}^{m}\sqcup f_{i}^{n})+\sum_{{\mathcal T},\mathcal{F}^{mn}} w(T_i)w(f_{m}^{m}\sqcup f_{j}^{n})\right]\\
&=P_{ji}+P_{mj}.
\end{align}
An identical argument holds for $P_{in}+P_{ji} = P_{jn}+P_{ij}$ \eqref{eq:struc_const_3}.\\

\noindent 3. \emph{Derivation of ${\mathcal P}_{mn}\le 1$\eqref{eq:struc_const_4}}:

From the definition of the structural coefficients \eqref{eq:struc_coeff}, we have
\begin{equation}
{\mathcal P}_{mn} = \left(\sqrt{w({\mathcal C}^{mn})}+\sqrt{w({\mathcal C}^{nm})}\right)^2 \frac{\sum_{\mathcal{F}^{mn}}w(f_{m}^{m}\sqcup f_{n}^{n})}{\sum_{\mathcal T}w(T_m)\sum_{\mathcal T}w(T_n)}.
\end{equation}
Next, we expand the denominator into sums of trees with the perturbed edge $T^{mn}\in{\mathcal T}^{mn}$ and those without $S^{mn}\in {\mathcal S}^{mn}$ (irrespective of orientation), and then apply the inequality of arithmetic and geometric means (AM-GM inequality),
\begin{align} \label{eq:bound_expree_P_mn}
   {\sum_{\mathcal T}w(T_m)\sum_{\mathcal T}w(T_n)} = & \left[\sum_{{\mathcal T}^{mn}}w\left(T^{mn}_{m}\right)+\sum_{{\mathcal S}^{mn}}w\left(S^{mn}_{m}\right)\right]\left[\sum_{{\mathcal T}^{mn}}w\left(T^{mn}_{n}\right)+\sum_{{\mathcal S}^{mn}}w\left(S^{mn}_{n}\right)\right] \\
   & \geq \left(\sqrt{\sum_{{\mathcal T}^{mn}}w\left(T^{mn}_{m}\right)\sum_{{\mathcal S}^{mn}}w\left(S^{mn}_{n}\right)}+\sqrt{\sum_{{\mathcal T}^{mn}}w\left(T^{mn}_{n}\right)\sum_{{\mathcal S}^{mn}}w\left(S^{mn}_{m}\right)}\right)^{2}.
\end{align}
Now, take the terms inside the first square root $w\left(T^{mn}_{m}\right)w\left(S^{mn}_{n}\right)$.
Because $T^{mn}_{m}$  is rooted at $m$, edge $e_{mn}$ linking $m$ and $n$ must indeed be oriented towards $m$.
Removal of this edge $e_{mn}$ from $T^{mn}_{m}$ forms a 2-forest, $f^m_m\sqcup f^n_n$.
If we then add that edge to $S^{mn}_{n}$, we form a cycle-graph $C^{mn}$.
Thus, $w\left(T^{mn}_{m}\right)w\left(S^{mn}_{n}\right)=w(C^{mn})w(f^m_m\sqcup f^n_n)$.
Applying this argument to every term above, we find
\begin{equation}
 {\sum_{\mathcal T}w(T_m)\sum_{\mathcal T}w(T_n)} \geq  \left(\sqrt{w({\mathcal C}^{mn})}+\sqrt{w({\mathcal C}^{nm})}\right)^2 \sum_{\mathcal{F}^{mn}}w(f_{m}^{m}\sqcup f_{n}^{n}),
\end{equation}
which implies the desired result.\\

The linear relationships between the structural coefficients implied by \eqref{eq:struc_const_1}-\eqref{eq:struc_const_4} confine the ${\mathcal P}_{ij}$  to a convex polytope, called the feasible polytope.  
For the three-state triangle graph we can visualize this polytope, see Fig.~\ref{fig:linear_optimization}.
\begin{figure}
     \centering
     \includegraphics[width=15.2cm]{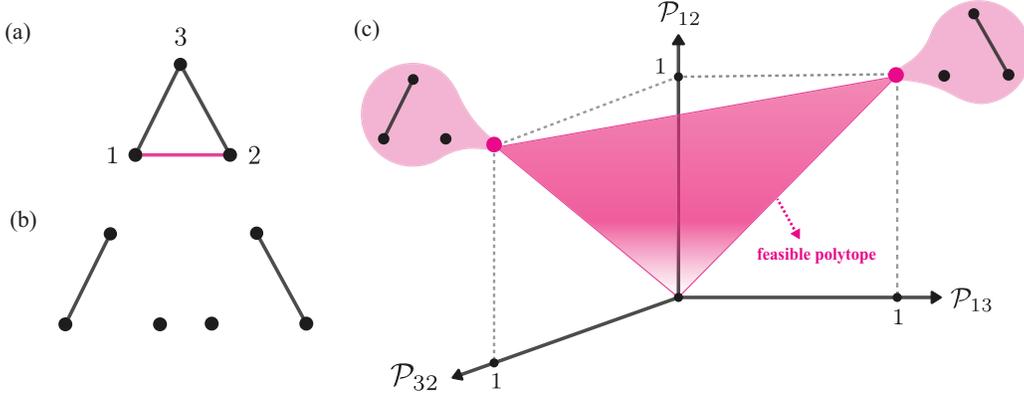}
     \caption{(a) Triangle graph with perturbed edge $\bar e_{12}$ (highlighted in pink).  (b) $2$-forests $\mathcal{F}^{12}$. (c) Feasible polytope: The only structural coefficients that are not trivially zero are $P_{12}$, $P_{13}$ and $P_{32}$.  As there is only one linearly independent equality constraint, $P_{12}=P_{13}+P_{32}$, the feasible polytope has dimension $d=2$. Its vertices are the trivial case $(0,0,0)$, as well as $(1,1,0)$ and $(1,0,1)$ that are each associated to a $2$-forest.}
     \label{fig:linear_optimization}
\end{figure}
Limits  to the response can then be deduced by identifying the values of the ${\mathcal P}_{ij}$ inside this polytope that maximize the response.
Namely, we can collect the constraints in \eqref{eq:struc_const_1}-\eqref{eq:struc_const_4} together to frame our question as the linear optimization problem: 
\begin{align}\label{eq:linOpt}
\underset{{\mathcal P}_{ij}}{\max} &\bigg| \sum_{i,j}\left(Q_j-Q_i\right)\pi_i\pi_j {\mathcal P}_{ij}\bigg| \quad{\rm such\ that\ for\ all\ } i\neq j \neq m,n \\
\label{eq:LP1}
&0\le {\mathcal P}_{ij}\le {\mathcal P}_{mj}\le {\mathcal P}_{mn}\le 1\\
\label{eq:LP2}
&0\le {\mathcal P}_{ij}\le {\mathcal P}_{in}\le {\mathcal P}_{mn}\le 1 \\
\label{eq:LP3}
& {\mathcal P}_{mn}={\mathcal P}_{mi}+{\mathcal P}_{in}\\
\label{eq:LP4}
&{\mathcal P}_{mi}+{\mathcal P}_{ij}= {\mathcal P}_{mj}+{\mathcal P}_{ji} \\
\label{eq:LP5}
&{\mathcal P}_{in}+{\mathcal P}_{ji}= {\mathcal P}_{jn}+{\mathcal P}_{ij}.
\end{align}
The fundamental theorem of linear programming then states that the solution to this optimization problem will be one of the vertices of the feasible polytope~\cite{Luenberger}.

To determine the vertices of the feasible polytope, we first determine its dimension $d$ given by the number of trivially nonzero structural coefficients less the number of linearly independent equality constraints:
\begin{equation}
\begin{split}
d &= \overbrace{1}^{{\mathcal P}_{mn}}+\overbrace{2(N-2)}^{{\mathcal P}_{mi},{\mathcal P}_{in}\ (i\neq m,n)}+\overbrace{(N-2)(N-3)}^{{\mathcal P}_{ij}\ (i\neq j\neq m,n)} - \overbrace{(N-2)}^{\rm(B34)}-\overbrace{(N-2)(N-3)/2}^{{\rm (B35)}, (i>j)}\\
&= 1+N-2 +(N-2)(N-3)/2
\end{split}
\end{equation}
A  vertex is then the unique solution to a subset of at least $d$ of the inequalities in \eqref{eq:LP1}-\eqref{eq:LP2} solved as equalities. 
In the following, we show that at these vertices all structural coefficients ${\mathcal P}_{ij}$ are either one or zero, that is they form a subset of the vertices of the positive unit hypercube.

We only need to focus on a collection of $d$ linearly independent structural coefficients, which we choose to be ${\mathcal P}_{mn}$, ${\mathcal P}_{mi}$ ($i\neq  m,n$), and ${\mathcal P}_{ij}$ ($i>j;\, i\neq j\neq m,n$).
We then express all the inequalities in \eqref{eq:LP1} and \eqref{eq:LP2} in terms of our independent variables.
After eliminating any redundant inequalities, we find for $i>j;\, i\neq j\neq m,n$,
\begin{align}
\label{eq:LP1_a}
&0\le {\mathcal P}_{ij}\le {\mathcal P}_{mj}\le {\mathcal P}_{mn}\le 1\\
\label{eq:LP2_a}
&{\mathcal P}_{mi}+{\mathcal P}_{ij}\le {\mathcal P}_{mn}\\
\label{eq:LP3_a}
&{\mathcal P}_{mj}\le{\mathcal P}_{mi}+ {\mathcal P}_{ij}
\end{align}
Clearly, if ${\mathcal P}_{mn}=0$, then all structural coefficients are zero, leading to a trivial solution that clearly cannot be a maximum.  
Thus, ${\mathcal P}_{mn}\neq 0$.

So we consider the case where ${\mathcal P}_{mn}\neq 0$.
We will now argue that at every vertex all the structural coefficients are either zero or one.
To determine the vertices we systematically choose one inequality to saturate and use that equality to fix a structural coefficient.
Doing that $d$ times leads to the conditions for a vertex, as long as the solution is consistent with all other inequalities.
We will begin by carrying out this program only for the string of inequalities in \eqref{eq:LP1_a}, identifying a collection of vertices on the unit hypercube.
Any other vertices can then be obtained by starting from a known vertex and then marching along every edge of the feasible polytope until arriving at another vertex, in the spirit of the simplex algorithm~\cite{Luenberger}.
It will turn out that starting from any known vertex, every edge connects it to another already-identified vertex on the unit hypercube.

First, we fix~${\mathcal P}_{mn}=1$, saturating one inequality.  
For each remaining structural coefficient we can fix its value by saturating either the upper or lower inequality in ~\eqref{eq:LP1_a}.
Specifically, we can choose either ${\mathcal P}_{mj}=1$ or ${\mathcal P}_{mj}={\mathcal P}_{i'j}$, for some $i'>j$.
If we fixed a particular structural coefficient via the second equality, ${\mathcal P}_{mj}={\mathcal P}_{i'j}$, then the only way to fix ${\mathcal P}_{i'j}$ using one of the remaining inequalities in \eqref{eq:LP1_a} is to set ${\mathcal P}_{mj}={\mathcal P}_{i'j}=0$.
This automatically sets all ${\mathcal P}_{mj}={\mathcal P}_{ij}=0$, for $i>j$.
On the other hand, if we had set ${\mathcal P}_{mj}=1$, then each ${\mathcal P}_{ij}$ for $i>j$ can be set to either ${\mathcal P}_{ij}=0$ or ${\mathcal P}_{ij}={\mathcal P}_{mj}=1$.
However, an inconsistency could arise with \eqref{eq:LP2_a}, if ${\mathcal P}_{ij}=1$ and ${\mathcal P}_{mi}=1$ ($2\le 1$).
So we must choose consistently  ${\mathcal P}_{mi}=0$ with ${\mathcal P}_{ij}=1$.
Similarly, an inconsistency could arise with \eqref{eq:LP3_a}, if ${\mathcal P}_{mj}=1$ and ${\mathcal P}_{ij}=0$ with ${\mathcal P}_{mi}=0$ ($1\le0$).
Thus, we must choose consistently ${\mathcal P}_{ij}=1$, when  ${\mathcal P}_{mj}=1$  and ${\mathcal P}_{mi}=0$.

We can characterize these vertices by recognizing that the condition that either ${\mathcal P}_{mj}=1$ or ${\mathcal P}_{mj}=0$ splits the set of states into two groups.
In the first group ${\mathcal P}_{mj}=1$, and $j$ is associated with the $f^n$ component.
Let us collect these nodes into a set $j\in V^n$.
The remaining set of nodes, ${\mathcal P}_{mi}=0$, can be characterized via \eqref{eq:LP3} by the condition ${\mathcal P}_{in}={\mathcal P}_{mn}-{\mathcal P}_{mi}=1$, which allows us to associate them to the $f^m$ component.
We call this set $i\in V^m$.
The remaining vertex conditions are consistent with this splitting.
Indeed, if $i\in V^m$, then ${\mathcal P}_{mi}=0$, implies ${\mathcal P}_{qi}=0$ for all $q$.
Similarly, $j\in V^n$, then ${\mathcal P}_{mj}=1$, implies either ${\mathcal P}_{qj}=1$ or ${\mathcal P}_{qj}=0$, depending on whether $q\in V^m$ or not.

The vertices we have so far identified turn out to be all the possible vertices. 
We can see this by using already identified vertices to detect any remaining ones.
Vertices are linked by edges of the polytope, where $d-1$ inequalities are saturated.
Thus, starting from any vertex we can identify additional vertices by marching along all incident edges.  
We accomplish this by taking one of the saturated equalities, and relax it by varying one of the coefficients.
This gives us the one degree of freedom required to delineate the edge.
To this end, let us choose a polytope vertex and label the states as $j_\alpha,j_\beta,\dots \in V^n$ and $i_\alpha,i_\beta,\dots\in V^m$.
Then one can check that nearly all the inequalities in \eqref{eq:LP1_a}-\eqref{eq:LP3_a} are saturated except 
\begin{align}
0={\mathcal P}_{j_\beta j_\alpha}\le {\mathcal P}_{mj_\alpha}=1, \qquad 0={\mathcal P}_{m,i_\alpha}+{\mathcal P}_{i_\beta,i_\alpha}\le {\mathcal P}_{mn}=1
\end{align}
These inequalities provide the only freedom for saturating new inequalities where we would find a new vertex. 
Thus, we can move any one of these coefficients from their current value to a new value that saturates either of these inequalities.
Though, at this new vertex it is clear again all structural coefficients will in fact be zero or one, as claimed.
Furthermore, as we move any coefficient from $1\to 0$ or $0\to 1$, the remaining equalities will keep all coefficients in the feasible polytope. 
The effect is just switching one of the states between the $V^n$ and $V^m$ sets.

The last step in characterizing the vertices of the polytope is to recognize that the splitting of the graph nodes into the sets $V^n$ and $V^m$ has to be consistent with the definition of the structural coefficients in terms of 2-forests. 
Take for example an $i\in V^m$ and $j\in V^n$, so that at this vertex ${\mathcal P}_{ij}=1$; however, this structural coefficient can only be nonzero if there exists at least one forest of the form $f^m_i\sqcup f^n_j$.
Thus, the only potential allowable choices of $V^n$ and $V^m$ is when they align with the vertex sets of a 2-forest, that is $V^n= V(f^n)$ and $V^m = V(f^m)$.\\

Now, with knowledge of the structural coefficients at the vertices, we can calculate the value of our objective function at these potential optima.
Recognizing that $V^m\sqcup V^n = V(G)$, so that $\sum_{j\in V^m}\pi_j+\sum_{j\in V^n}\pi_j=1$, we have
\begin{equation}
\begin{split}
\sum_{i,j}\left(Q_j-Q_i\right)\pi_i\pi_j {\mathcal P}_{ij}\bigg|_{{\rm vertex}} &= \sum_{i\in V^m}\sum_{j\in V^n}\left(Q_j-Q_i\right)\pi_i\pi_j \\
& = \sum_{j\in V^n}Q_j\pi_j\sum_{i\in V^m}\pi_i-\left(\langle Q\rangle -\sum_{j\in V^n}Q_j \pi_j\right) \sum_{j\in V^n}\pi_j\\
& =  \sum_{j\in V^n}Q_j\pi_j-\langle Q\rangle\sum_{j\in V^n}\pi_j\\
& =\llangle Q,\delta(V^n)\rrangle=-\llangle Q,\delta(V^m)\rrangle,
\end{split}
\end{equation} 
with $\delta_i(V)$ the characteristic function of a vertex set $V$, taking value one when $i\in V$ and zero otherwise. We arrive at our bound \eqref{eq:struc_bound} by noting that any one of the vertices could be the maximum.
The particular one depends on the observable and the steady-state distribution.

\subsection{Optimal network topologies}

The conditions on the rates for saturation of~\eqref{eq:bound_B} suggest design principles for constructing optimal network topologies that maximize the response under thermodynamic and noise constraints.  
There are two key bounds that must be saturated.
The first is the limit on the response magnitude ${\mathcal M}$ \eqref{eq:responseMagnitudeBound}, and the second is the linear optimization problem that is setup in \eqref{eq:linOpt}.
We address each in turn.

The bound on ${\mathcal M}$ \eqref{eq:responseMagnitudeBound} becomes an equality when every cycle through $m$ and $n$ has the same weight.  
The simplest and perhaps most generic situation where this occurs is when there is a single cycle $C^{mn}$ through $m$ and $n$.
Thus, to saturate the inequality we cut all cycles passing through $m$ and $n$ except for one.
To cut a cycle, we must delete one edge in that cycle, by sending the rates on that edge to zero.

The second condition on an optimal topology emerges from the optimal solution of the linear optimization problem in \eqref{eq:linOpt}.
Our derivation reveals that at the optimum the structural coefficients ${\mathcal P}_{ij}$ \eqref{eq:struc_coeff} are either zero or one.
The distinction depends on whether they contain the weight of a forest which aligns with the optimal vertex set.
The simplest scenario where this occurs is where there is a single dominant $2$-forest. 
This can be arranged by setting all the rates on the $2$-forest to be large, which in effect makes every structural coefficient containing that $2$-forest approximately one (when there is a single cycle), and all others are zero.
Fast rates contract edges of the network, replacing the source and target of the edge with a single effective node.  

Putting these observations together suggests that the optimal topology is composed of two fast islands formed by the trees contained in a single forest, which are each contracted into a pair of single nodes.  These two islands are then linked up by a pair of slow edges that complete the single cycle network.

\section{Multiple connected edges symmetric perturbation}
\label{sec:mutli-symmetric}

Here, we expand on the previous bounds to include the symmetric perturbation of multiple connected edges of the graph, and derive~\eqref{eq:bound_B_multi}. 
Our assumption is that these edges (with their incident nodes) form a subgraph $H^{mn}$ of $G$ that only connects to the rest of the graph at two vertices, which we call $m$ and $n$, as illustrated in Fig.~\ref{fig:set_decomposition}.
Note that $m$ and $n$ may not be directly linked by any single edge, but there is at least one path between $m$ and $n$ through $H^{mn}$.
All the edges not in $H^{mn}$ and incident to either $m$ or $n$ form a cut-set of $G$, which we label as $K^{mn}$---removing all of them splits $G$ into two separate connected components.
\begin{figure}
     \centering
     \includegraphics[width=15.2cm]{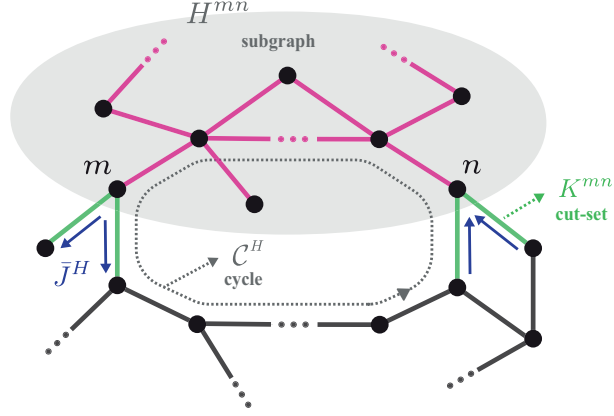}
     \caption{Multi-edge perturbation in a generic graph: all edges (pink) in the subgraph $H^{mn}$ (gray) are perturbed uniformly.  Removing all edges of the cut-set $K^{mn}$ (green) splits the graph into two separate subgraphs. The probability current $\bar J ^H$ (blue arrows) flowing in and out of $H^{mn}$ is due to cycles $\mathcal C^H$ that straddle the perturbed and unperturbed regions.}
     \label{fig:set_decomposition}
\end{figure}

To bound the response of an observable to a concerted and uniform symmetric perturbation of all edges in $H^{mn}$, we follow closely our derivation for a single edge and consider the derivative
\begin{equation}\label{eq:obs_res_multi}
\sum_{{\bar e}_{lk} \in H^{mn}}\frac{\partial \langle Q\rangle }{\partial B_{lk}} =\sum_{j=1}^N Q_j \sum_{{\bar e}_{lk} \in H^{mn}} \frac{\partial \pi_j}{\partial B_{lk}},
\end{equation}
with the steady-state responses solving the set of inhomogenous linear equations
\begin{equation}\label{eq:sum_pert_multi}
    \sum_{j=1}^N W_{ij} \left(\sum_{{\bar e}_{lk}\in H^{mn}}\frac{\partial \pi_j}{\partial B_{lk}}\right)= \sum_{{\bar e}_{lk}\in H^{mn}} J_{lk}\left(\delta_{il}-\delta_{ik}\right).
\end{equation}
At this point, we recall that conservation of probability requires that the total probability current flowing into or out of any vertex is zero, $\sum_i J_{ij}=\sum_j J_{ij}=0$. 
Therefore the sums in \eqref{eq:sum_pert_multi} at vertices internal to $H^{mn}$ cancel.
Only at $m$ and $n$ does the sum on probability currents leave an imbalance equal to the total probability flowing in or out of the subgraph $H^{mn}$,
\begin{equation}\label{eq:sum_pert_multi_2}
    \sum_{j=1}^N W_{ij} \left(\sum_{e_{lk}\in H^{mn}}\frac{\partial \pi_j}{\partial B_{lk}}\right) = {\bar J}^{H}(\delta_{im}-\delta_{in}),
\end{equation}
with total probability flow out of $H^{mn}$ (which is the same as into due to probability conservation) given by the sum of the currents on all edges in the cut-set oriented away from $m$  
\begin{equation}
{\bar J}^{H}=\sum_{ e_{im}\in K^{mn}} {\bar J}_{im} = \frac{1}{\mathcal N} \left[w\left({\mathcal C}^{H}\right)-w\big(\tilde{\mathcal C}^{H}\big)\right].
\end{equation}
Here, we have further observed that this probability current is only to due to cycle-graphs in ${\mathcal C}^{H}$ whose cycles straddle $H^{mn}$ and its complement, oriented such that they exit at $m$ and enter at $n$, as illustrated in Fig.~\ref{fig:set_decomposition}. 
The solution of \eqref{eq:sum_pert_multi_2} can be organized using 2-forests (cf.~\eqref{eq:solution_forest}),
\begin{equation}
  \sum_{{\bar e}_{lk}\in H^{mn}}\frac{\partial \pi_j}{\partial B_{lk}} =\frac{{\bar J}^{H}}{\mathcal{N}}\sum_{\mathcal{F}^{mn},j}w\left(f_{j}^{m}\sqcup f_{i}^{n}\right)-w\left(f_{i}^{m}\sqcup f_{j}^{n}\right),
\end{equation}
Substitution into \eqref{eq:obs_res_multi} and rearranging as before, leads to
\begin{equation}
   \left| \sum_{{\bar e}_{lk} \in H^{mn}}\frac{\partial \langle Q\rangle }{\partial B_{lk}} \right|={\mathcal M}^{H} \bigg| \sum_{i,j}\left(Q_j-Q_i\right)\pi_i\pi_j {\mathcal P}^{H}_{ij}\bigg|,
\end{equation}
with overall magnitude
\begin{equation}
{\mathcal M}^{H} = \frac{ \left|w({\mathcal C}^{H})-w(\tilde{\mathcal C}^{H})\right|}{\left(\sqrt{w({\mathcal C}^{H})}+\sqrt{w(\tilde{\mathcal C}^{H})}\right)^2},
\end{equation}
and structural coefficients
\begin{equation}
{\mathcal P}^{H}_{ij} = \left(\sqrt{w({\mathcal C}^{H})}+\sqrt{w(\tilde{\mathcal C}^{H})}\right)^2  P_{ij},\qquad{\rm where}\qquad  P_{ij} = \frac{\sum_{\mathcal{F}^{mn}}w(f_{i}^{m}\sqcup f_{j}^{n})}{\sum_{\mathcal T}w(T_i)\sum_{\mathcal T}w(T_j)}.
\end{equation}
In the following, we bound the magnitude as well as the sum over structural coefficients.

The bound on ${\mathcal M}^{H}$ follows the exact same line of reasoning we used to bound ${\mathcal M}$ in Sec.~\ref{sec:mag}. 
The result is
\begin{equation}
{\mathcal M}^{H}\le \tanh({\mathcal F}_{\rm max}/4),
\end{equation}
except here ${\mathcal F}_{\rm max} = \max_{C\in {\mathcal C}^{H}} |F(C)| $ as it emerges from ratios of elements of ${\mathcal C}^{H}$ and $\tilde{\mathcal C}^{H}$.

Next, we observe that the structural coefficients ${\mathcal P}^{H}_{ij}$ are defined in terms of the same type of forest-tree ratios that appeared in our analysis of a single edge perturbation.
Furthermore, the linear relationships between the forest-tree ratios in \eqref{eq:struc_const_1}-\eqref{eq:struc_const_3} did not depend on the vertices being linked by an edge, and thus hold here as well.  
The only potential difference from our single-edge analysis is the bound ${\mathcal P}^{H}_{mn}\le 1$ (cf.~\eqref{eq:struc_const_4}), but this holds as well.
From its definition, we have 
\begin{equation}
{\mathcal P}^{H}_{mn} = \left(\sqrt{w({\mathcal C}^{H})}+\sqrt{w(\tilde{\mathcal C}^{H})}\right)^2  \frac{\sum_{\mathcal{F}^{mn}}w(f_{m}^{m}\sqcup f_{n}^{n})}{\sum_{\mathcal T}w(T_m)\sum_{\mathcal T}w(T_n)}.
\end{equation}
We lower bound the denominator by observing that since $H^{mn}$ is only incident to the rest of the graph at two nodes, any path in a tree that links $m$ and $n$ must be contained entirely in $H^{mn}$ or its complement; in other words no path can enter or leave $H^{mn}$ without crossing through $m$ or $n$.
This allows us to divide the sum over trees in the denominator into those $T^{n {\buildrel H\over \leftrightsquigarrow} m}$ where the unique path that connects $m$ and $n$ is contained entirely in $H^{mn}$ and those $S^{n {\buildrel H\over \leftrightsquigarrow} m}\in {\mathcal S}^{n {\buildrel H\over \leftrightsquigarrow} m}$ where the path is entirely contained in the complement of $H^{mn}$:
\begin{equation} 
\begin{split}
 \sum_{\mathcal T}&w(T_m)\sum_{\mathcal T}w(T_n)  \\
  & = \left[\sum_{{\mathcal T}^{n {\buildrel H\over \leftrightsquigarrow} m}}w\left(T_m^{n {\buildrel H\over \leftrightsquigarrow} m}\right)+\sum_{{\mathcal S}^{n {\buildrel H\over \leftrightsquigarrow} m}}w\left(S_n^{m {\buildrel H\over \leftrightsquigarrow} m}\right)\right]
  \left[\sum_{{\mathcal T}^{n {\buildrel H\over \leftrightsquigarrow} m}}w\left(T_n^{n {\buildrel H\over \leftrightsquigarrow} m}\right)+\sum_{{\mathcal S}^{n {\buildrel H\over \leftrightsquigarrow} m}}w\left(S_n^{n {\buildrel H\over \leftrightsquigarrow} m}\right)\right].
  \end{split}
  \end{equation}
  Expanding the denominator and applying the AM-GM inequality, we have
  \begin{align}\label{eq:mutli_denominator_bound}
    \sum_{\mathcal T}w(T_m)\sum_{\mathcal T}w(T_n) \geq 
    \left(
    \sqrt{\sum_{{\mathcal T}^{n {\buildrel H\over \leftrightsquigarrow} m},{\mathcal S}^{n {\buildrel H\over \leftrightsquigarrow} m}}w\left(T_m^{n {\buildrel H\over \leftrightsquigarrow} m}\right)w\left(S_n^{n {\buildrel H\over \leftrightsquigarrow} m}\right)}
    +\sqrt{\sum_{{\mathcal T}^{n {\buildrel H\over \leftrightsquigarrow} m},{\mathcal S}^{n {\buildrel H\over \leftrightsquigarrow} m}}w\left(T_n^{n {\buildrel H\over \leftrightsquigarrow} m}\right)w\left(S_n^{m {\buildrel H\over \leftrightsquigarrow} m}\right)}
    \right)^{2}.
\end{align}
Now, take the terms inside the first square root $w\left(T_m^{n {\buildrel H\over \leftrightsquigarrow} m}\right)w\left(S_n^{n {\buildrel H\over \leftrightsquigarrow} m}\right)$.
The part of the tree $T_m^{n {\buildrel H\over \leftrightsquigarrow} m}$ inside the perturbed region $H^{mn}$ is by definition connected and rooted at $m$.
By contrast, the part of the tree $S_n^{n {\buildrel H\over \leftrightsquigarrow} m}$ in $H^{mn}$ is not connected, but is composed of two connected components, one rooted at $m$ and the other rooted at $n$.
This is the only arrangement possible in a tree rooted at $n$ if $m$ is to be linked to $n$ in the complement of $H^{mn}$.
We now swap all the perturbed edges in $T_m^{n {\buildrel H\over \leftrightsquigarrow} m}$ with all the perturbed edges in $S_n^{n {\buildrel H\over \leftrightsquigarrow} m}$.
The tree that initially had the path between $m$ and $n$ in $H^{mn}$ is now disconnected and forms a 2-forest $f^m_m\sqcup f^n_n$.
The tree that did not have a path between $m$ and $n$ in $H^{mn}$ has one now, forming a cycle-graph in the set ${\mathcal C}^{H}$.
This is just like the single-edge perturbation, except instead of swapping a single edge, we swap all the perturbed edges together.
Applying a similar argument to the second sum in  \eqref{eq:mutli_denominator_bound}, results in
\begin{equation}
 {\sum_{\mathcal T}w(T_m)\sum_{\mathcal T}w(T_n)} \geq  \left(\sqrt{w({\mathcal C}^{H})}+\sqrt{w(\tilde{\mathcal C}^{H})}\right)^2 \sum_{\mathcal{F}^{mn}}w(f_{m}^{m}\sqcup f_{n}^{n}),
\end{equation}
which implies the desired result.

\section{Arbitrary single rate perturbation}
\label{sec:single_rate}

In this section, we adapt the methods developed above to derive~\eqref{eq:W_perturb}. 
The response of an observable to the logarithmic perturbation of a single kinetic rate $W_{mn}$ is
\begin{equation}\label{eq:W_pert}
\frac{\partial \langle Q\rangle}{\partial \ln W_{mn}} = \sum_{j=1}^N Q_j\frac{\partial \pi_j}{\partial \ln W_{mn}}.
\end{equation}
Here, the master equation implies the responses of the steady-state  distribution satisfy
\begin{equation}
\sum_{j=1}^N W_{ij}  \frac{\partial\pi_{j}}{\partial \ln W_{mn}} = W_{mn}\pi_n(\delta_{in}-\delta_{im}).
\end{equation}
The solution can be compactly organized in terms of 2-forests \eqref{eq:solution_forest} as 
\begin{equation}
    \frac{\partial\pi_{i}}{\partial \ln W_{mn}}=\frac{W_{mn}\pi_n}{\mathcal{N}}\sum_{\mathcal{F}^{mn},j}w\left(f_{i}^{m}\sqcup f_{j}^{n}\right)-w\left(f_{j}^{m}\sqcup f_{i}^{n}\right),
\end{equation}
Substituting into \eqref{eq:W_pert} and reorganizing, we recover the structure in Appendix~\ref{sec:symmetric},
\begin{equation}
   \left| \frac{\partial\langle Q\rangle}{\partial \ln W_{mn}}\right|=\bigg| \sum_{i,j}\left(Q_j-Q_i\right)\pi_i\pi_j {\mathcal P}^W_{ij}\bigg|,
\end{equation}
where in this case the structural coefficients are 
\begin{equation}
{\mathcal P}^W_{ij} =W_{mn}\sum_{\mathcal T}w(T_n) P_{ij},\qquad{\rm with}\qquad P_{ij} = \frac{\sum_{\mathcal{F}^{mn}}w(f_{i}^{m}\sqcup f_{j}^{n})}{\sum_{\mathcal T}w(T_i)\sum_{\mathcal T}w(T_j)}.
\end{equation}

We again conclude that the structural coefficients retain the linear relationships  presented in \eqref{eq:struc_const_1}-\eqref{eq:struc_const_4} necessary to bound this sum.
The only potential change is the overall magnitude, requiring us to demonstrate that ${\mathcal P}^W_{mn}\le 1$ as in  \eqref{eq:struc_const_4}.
From the definition of the structural coefficients, we have
\begin{equation}
{\mathcal P}^W_{mn} =W_{mn}\sum_{\mathcal T}w(T_n)\frac{\sum_{\mathcal{F}^{mn}}w(f_{m}^{m}\sqcup f_{n}^{n})}{\sum_{\mathcal T}w(T_m)\sum_{\mathcal T}w(T_n)}=\frac{\sum_{\mathcal{F}^{mn}}W_{mn}w(f_{m}^{m}\sqcup f_{n}^{n})}{\sum_{\mathcal T}w(T_m)}.
\end{equation}
Recognizing that addition of the edge $e_{mn}$ to a 2-forest $f^m_m\sqcup f^n_n$ results in a tree $T_m^{n\to m}$ rooted at $m$ with the edge directed from $n\to m$, so that $W_{mn}w(f_{m}^{m}\sqcup f_{n}^{n}) = w(T^{n\to m}_m)$, we have, 
\begin{equation}
{\mathcal P}^W_{mn} =\frac{\sum_{{\mathcal T}}w(T^{n\to m}_m)}{\sum_{\mathcal T}w(T_m)}\le 1,
\end{equation}
where the inequality follows because the set of all trees includes more trees then just those that contain the edge $e_{mn}$.

\section{Operational limits: bounds on covariance}
\label{sec:opt_limits}

To bound the recurring covariance $\underset{\mathcal{V}^{mn}}{\max} \Big|\left\llangle Q,\delta(V^{m})\right\rrangle\Big|$ we start by noting that each $\delta(V^m)$ is nonnegative and bounded by one ($0\le \delta(V^m)\le 1$). Let us consider the set of all such bounded observables, ${\mathcal BO} = \{A| 0\le A_i\le 1\}$, of which $\delta(V^m)$ is a member. Then, the correlation we wish to constrain can be bounded by the maximum over all steady-states $\pi_i$ and all bounded observables, keeping the average $\langle Q\rangle$ and variance $\llangle Q^2\rrangle$ fixed:
\begin{equation}
\underset{\mathcal{V}^{mn}}{\max} \Big|\left\llangle Q,\delta(V^{m})\right\rrangle\Big| \le \underset{\{\pi,A\in {\mathcal BO}|\langle Q\rangle,\llangle Q^2\rrangle\}}{\max} \Big|\left\llangle Q,A\right\rrangle\Big|.
\end{equation}
To facilitate this calculation, we use that $\langle Q\rangle$ is fixed to shift the observable so that its mean is zero,
\begin{equation}
{\tilde Q}_i = Q_i-\langle Q\rangle,
\end{equation}
and introduce a notation for the value of the fixed variance $\llangle Q^2\rrangle =\langle {\tilde Q}^2\rangle= v$.
As the covariance is invariant to constant shifts, our problem becomes
\begin{equation}\label{eq:cov_1}
\underset{\{\pi,A\in {\mathcal BO}|\langle Q\rangle,\llangle Q^2\rrangle\}}{\max} \Big|\llangle Q,A\rrangle\Big|=\underset{\{\pi,A\in {\mathcal BO}|\langle {\tilde Q}\rangle, \langle {\tilde Q}^2\rangle\}}{\max} \Big|\llangle {\tilde Q},A\rrangle\Big| =\underset{\{\pi,A\in {\mathcal BO}|\langle{\tilde Q}\rangle, \langle {\tilde Q}^2\rangle\}}{\max} \left|\sum_{j=1}^N {\tilde Q}_j A_j \pi_j\right|
\end{equation}

We first perform the maximization over all $A\in {\mathcal BO}$.
As $\langle {\tilde Q}\rangle =0$, we can divide the vertices into those where the observable is positive ${\tilde Q}^+ = \{ i | {\tilde Q}_i \ge 0\}$, and those where it is negative  ${\tilde Q}^- = \{ i | {\tilde Q}_i < 0\}$.
We can clearly maximize \eqref{eq:cov_1} by only keeping terms in the sum that have the same sign.  
If we keep only positive terms, by setting $A_i=1$ for all $i\in {\tilde Q}^+$ and $A_i=0$ otherwise, we find
\begin{equation}
\underset{\{\pi,A\in {\mathcal BO}|\langle Q\rangle,\llangle Q^2\rrangle\}}{\max} \Big|\llangle Q,A\rrangle\Big|= \underset{\{\pi |\langle {\tilde Q}\rangle, \langle {\tilde Q}^2\rangle\}}{\max} \left|\sum_{j\in{\tilde Q}^+}  {\tilde Q}_j \pi_j\right|.
\end{equation}
Notice if we had kept only negative terms, we would have arrived at the same bound as $\left|\sum_{j\in{\tilde Q}^+}  {\tilde Q}_j \pi_j\right| = \left|\sum_{j\in{\tilde Q}^-}  {\tilde Q}_j \pi_j\right|$.

We have thus reduced our analysis to the following  linear optimization problem:
\begin{align}
 \underset{\pi}{\max}  &\left|\sum_{j\in{\tilde Q}^+}  {\tilde Q}_j \pi_j\right|\quad{\rm such\ that}\\
 \label{eq:cov_opt_1}
&0\le \pi_i \le 1\\
 \label{eq:cov_opt_2}
&\sum_i\pi_i=1 \\
 \label{eq:cov_opt_3}
&\sum_i{\tilde Q}_i \pi_i=0\\
 \label{eq:cov_opt_4}
&\sum_i{\tilde Q}^2_i \pi_i=v.
\end{align}
The potential maxima are given by the vertices of the convex polytope defined by the above constraints.
With three equality constraints the dimension of this polytope is $d=N-3$.
Thus, at a vertex we need to saturate $d$ of the inequalities in \eqref{eq:cov_opt_1}.
We can only do this by setting $d$ of the $\pi_i$'s to either 1 or 0; however, if any $\pi_i=1$, then probability conservation \eqref{eq:cov_opt_2} requires the probability at all other sites to be zero, which is a situation where we cannot maintain the mean constraint on ${\tilde Q}$ \eqref{eq:cov_opt_3}.
As a result, vertices are characterized by the steady-state distribution having only three nonzero elements and the rest zero.
To have $\sum_i{\tilde Q}_i \pi_i=0$, one of the nonzero elements of $\pi$ must be in ${\tilde Q}^+$, let us call this state $i=+$ and have value ${\tilde Q}_+$; another must be in the negative region ${\tilde Q}^-$, let us call it $i=-$ with value $-{\tilde Q}_-$; the third state we will call $i=o$ and we will take it to have a value in between $-{\tilde Q}_-\le {\tilde Q}_o\le {\tilde Q}_+$.
The probabilities at these sites, $\pi_+$, $\pi_o$ and $\pi_-$, are then determined by the equality constraints \eqref{eq:cov_opt_2} - \eqref{eq:cov_opt_4}, which read
\begin{equation}
\left(\begin{array}{ccc}
1 & 1 & 1 \\
{\tilde Q}_+ & {\tilde Q}_o & -{\tilde Q}_- \\
{\tilde Q}_+^2 & {\tilde Q}_o^2 & {\tilde Q}_-^2
\end{array}\right)
\left(\begin{array}{c}
\pi_+ \\
\pi_o \\
\pi_-
\end{array}\right)
=
\left(\begin{array}{c}1 \\0 \\ v\end{array}\right).
\end{equation}
The unique solution is
\begin{equation}
\pi_+=\frac{v-{\tilde Q}_o{\tilde Q}_-}{({\tilde Q}_+-{\tilde Q}_0)({\tilde Q}_++{\tilde Q}_-)},\qquad \pi_o=\frac{{\tilde Q}_+{\tilde Q}_- - v}{({\tilde Q}_+-{\tilde Q}_0)({\tilde Q}_0+{\tilde Q}_-)}, \qquad \pi_-=\frac{v+{\tilde Q}_+{\tilde Q}_0}{({\tilde Q}_0+{\tilde Q}_-)({\tilde Q}_++{\tilde Q}_-)}.
\end{equation}
Positivity of the steady-state probabilities further requires the constraints
\begin{equation}\label{eq:obs_const}
v\ge {\tilde Q}_o{\tilde Q}_-,\qquad {\tilde Q}_+{\tilde Q}_- \ge  v, \qquad v\ge -{\tilde Q}_+{\tilde Q}_o.
\end{equation}

The value of our objective function at a vertex then depends on whether ${\tilde Q}_o$ is positive or negative:
\begin{align}\label{eq:var_vertex1}
&\left|\sum_{j\in{\tilde Q}^+}  {\tilde Q}_j \pi_j\right|_{{\rm vertex}, {\tilde Q}_0\le 0} = \left|{\tilde Q}_+\pi_+\right| = \left|\frac{{\tilde Q}_+(v-{\tilde Q}_o{\tilde Q}_-)}{({\tilde Q}_+-{\tilde Q}_0)({\tilde Q}_++{\tilde Q}_-)}\right|\\
\label{eq:var_vertex2}
&\left|\sum_{j\in{\tilde Q}^+}  {\tilde Q}_j \pi_j\right|_{{\rm vertex}, {\tilde Q}_0\ge 0} = \left|{\tilde Q}_+\pi_++{\tilde Q}_o\pi_o\right| = \left|\frac{{\tilde Q}_-(v+{\tilde Q}_+{\tilde Q}_0)}{({\tilde Q}_0+{\tilde Q}_-)({\tilde Q}_++{\tilde Q}_-)}\right|.
\end{align}
To find the vertex with the largest value, we next have to maximize over the values of the observable. We will analyze the case ${\tilde Q}_o\ge 0$, and as it turns out, the maximum is the same when ${\tilde Q}_o\le 0$.
We will allow ${\tilde Q}_+$, ${\tilde Q}_o$, and ${\tilde Q}_-$ to vary over all real numbers between the observable's maximum and minimum value, even if such values are not attained at any particular state.
Thus, we now have the bound on the maximum
\begin{equation}
 \underset{\pi}{\max}  \left|\sum_{j\in{\tilde Q}^+}  {\tilde Q}_j \pi_j\right| \le  \max_{{\tilde Q}_+,{\tilde Q}_-,{\tilde Q}_o}\left|\frac{{\tilde Q}_-(v+{\tilde Q}_+{\tilde Q}_0)}{({\tilde Q}_0+{\tilde Q}_-)({\tilde Q}_++{\tilde Q}_-)}\right|.
 \end{equation}
Observe that  that is a monotonically increasing function of ${\tilde Q}_o$, whose value is limited due to \eqref{eq:obs_const} by the constraints ${\tilde Q}_o \le v/{\tilde Q}_-\le {\tilde Q}_+$.
Thus,  the maximum is attained when the value of the observable takes is largest value ${\tilde Q}_o = v/{\tilde Q}_-$,
\begin{equation}\label{eq:var_qMinus}
 \underset{\pi}{\max}  \left|\sum_{j\in{\tilde Q}^+}  {\tilde Q}_j \pi_j\right| \le \max_{{\tilde Q}_-}\left|\frac{{\tilde Q}_- v}{{\tilde Q}_-^2+v}\right|.
 \end{equation}
 Notice that ${\tilde Q}_+$ has dropped from the calculation, because when ${\tilde Q}_o = v/{\tilde Q}_-$ there is no probability on $\pi_+ =0$, and the distribution is now peaked at two sites: $\pi_o={\tilde Q}^2_-/({\tilde Q}^2_-+v)$ and $\pi_-=v/({\tilde Q}^2_-+v)$.
 Now, \eqref{eq:var_qMinus} has a local maximum of $\sqrt{v}/2$ when ${\tilde Q}_-=\sqrt{v}$, and this is the maximum as long $\sqrt{v}$ is a value obtainable by the observable.
 However,  ${\tilde Q}_-$ is constrained by the observable's smallest (or most negative) value ${\tilde Q}_{m} = Q_m-\langle Q\rangle$ and its largest value ${\tilde Q}_{M} = Q_M-\langle Q\rangle$ via $v/{\tilde Q}_M\le v/{\tilde Q}_o={\tilde Q}_- \le | {\tilde Q}_m|$.
This leads to three possibilities:
(1) $\sqrt{v}\le {\tilde Q}_m, {\tilde Q}_M$ lies in the domain of ${\tilde Q}_-$ and the maximum is $\sqrt{v}/2$.
 (2) If ${\tilde Q}_m\le {\tilde Q}_M$, then we can have $|{\tilde Q}_m| \le  \sqrt{v} \le {\tilde Q}_M$.  In this case, the maximum is obtained on the boundary where ${\tilde Q}_- = |{\tilde Q}_m|$ with value $|{\tilde Q}_m|v/({\tilde Q}_m^2+v)$.
 (3) If ${\tilde Q}_M\le {\tilde Q}_m$, then we can have $|{\tilde Q}_M| \le  \sqrt{v} \le {\tilde Q}_m$.  In this case, the maximum is obtained on the boundary where ${\tilde Q}_- = {\tilde Q}_M$ with value ${\tilde Q}_Mv/({\tilde Q}_M^2+v)$.
 Finally, the largest value that $v$ can attain occurs when the domain for ${\tilde Q}_-$ shrinks to nothing, which occurs when $v={\tilde Q}_M  |{\tilde Q}_m|$.
 Collecting these observation leads to the expression in \eqref{eq:operational3}.

\section{Rate constants and observable values for Figures 6 and 7}
\label{sec:fig_values}

The rates that correspond to reaching the optimal network topology for the receptor binding model in Fig.~\ref{fig:model} are listed in Table~\ref{tab:rates_fig6}. The values of the rates are chosen to fix the thermodynamic driving $\Delta\mu/kT= \{1,5\}$. 
A multiplicative factor is included in the rates internal to the islands, $\varepsilon=10^4$, in order to impose a timescale separation between transitions internal to the islands relative to between islands. 
\begin{table}[h]
\caption{\label{tab:rates_fig6} Rate constants for the optimal network topology in Fig.~\ref{fig:receptorBound}.}
\begin{ruledtabular}
\begin{tabular}{c c c}
     & $\Delta\mu/kT = 1$ & $\Delta\mu/kT = 5$ \\
    \hline
    $k_1$ & $0.457986$ & $0.457986$ \\
    $\tilde k_1$ & $5.09981$  & $37.6828$  \\
    $k_2$ & $0.457986 \, \varepsilon$  & $0.457986 \,\varepsilon$  \\    
    $\tilde k_2$ & $1.20478 \, \varepsilon$  &  $1.20478 \, \varepsilon$ \\
    $k_3$ & $3.09319 \, \varepsilon$ & $3.09319 \, \varepsilon$  \\
    $\tilde k_3$ & $0.251558 \, \varepsilon$ & $0.251558 \, \varepsilon$ \\
    $k_4$ & $1.98634$ & $14.6772$ \\
    $\tilde k_4$ & $14.6772$ & $14.6772$ \\
\end{tabular}
\end{ruledtabular}
\end{table}

The three purple points in Fig.~\ref{fig:hill_curves}(a) that correspond to the purple binding curves in Fig.~\ref{fig:hill_curves}(b) were generated using the rate constants and observable values in Table~\ref{tab:rates}. 
The values are chosen so that the cycle force is equal to $2$ in all cases.
Note that in Fig.~\ref{fig:hill_curves}(b) the value of $c$ for which $\langle f \rangle_c$ is equal to $0.5$ depends on the rate constants. In the interest of creating a clear visualization, we numerically find these values, which we denote as $K$ and plot each curve as a function of the normalized concentration $c/K$. 
\begin{table}[h]
\caption{\label{tab:rates} Rate constants and the observable values for the purple points $1$, $2$ and $3$ in Fig.~\ref{fig:hill_curves}(a).}
\begin{ruledtabular}
\begin{tabular}{c c c c}
     & Point $1$ & Point $2$ & Point $3$ \\
    \hline
    $k_1$ & 27.7025 & 12.530555 & 0.001309 \\
    $\tilde k_1$ & 0.467072  & 10.464932  & 1.789962 \\
    $k_2$ & 0.664276 & 29.406176 & 0.236687 \\    
    $\tilde k_2$ & 8.4392  & 57.100834 & 272.941976 \\
    $k_3$ & 1.97131 & 320.175943 & 76.741704 \\
    $\tilde k_3$ & 0.142501 & 12.530555 & 0.023359\\
    $k_4$ & 4.07565 & 516.106104 & 6.817252 \\
    $\tilde k_4$ & 0.00288908 & 516.106104 & 0.018179 \\
    $f_2=f_4$ & 0.571365 & 0.559462 & 0.416399\\
\end{tabular}
\end{ruledtabular}
\end{table}

\end{widetext}

\bibliography{FDT.bib,PhysicsTexts.bib}

\end{document}